\documentclass[letterpaper, preprint, paper,11pt]{AAS}	

\usepackage{bm}
\usepackage{amsmath}
\usepackage{subcaption}
\usepackage[colorlinks=true, pdfstartview=FitV, linkcolor=black, citecolor= black, urlcolor= black]{hyperref}
\usepackage{overcite}
\usepackage{footnpag}			      	

\usepackage{amsmath}
\usepackage{amsfonts}
\usepackage{todonotes}
\usepackage{booktabs}
\usepackage{algorithm}
\usepackage{algpseudocode}
\usepackage{amssymb}

\PaperNumber{23-271}

\begin{document}

\title{
Sensitivity Analysis of Separation Time along Weak Stability Boundary Transfers
}

\author{Isabel Nolton\thanks{Undergraduate student, School of Aerospace Engineering, Georgia Institute of Technology, GA 30332, USA},  
Kento Tomita\thanks{PhD Candidate, School of Aerospace Engineering, Georgia Institute of Technology, GA 30332, USA},
Yuri Shimane\thanks{PhD Candidate, School of Aerospace Engineering, Georgia Institute of Technology, GA 30332, USA},
Koki Ho\thanks{Associate Professor, School of Aerospace Engineering, Georgia Institute of Technology, GA 30332, USA}
}

\maketitle{}

\begin{abstract}
This study analyzes the sensitivity of the dynamics around Weak Stability Boundary Transfers (WSBT) in the elliptical restricted three-body problem. With WSBTs increasing popularity for cislunar transfers, understanding its inherently chaotic dynamics becomes pivotal for guiding and navigating cooperative spacecrafts as well as detecting non-cooperative objects. We introduce the notion of separation time to gauge the deviation of a point near a nominal WSBT from the trajectory’s vicinity. Employing the Cauchy-Green tensor to identify stretching directions in position and velocity, the separation time, along with the Finite-Time Lyapunov Exponent are studied within a ball of state uncertainty scaled to typical orbit determination performances.

\end{abstract}




\section{Introduction}
Steering the motion through a chaotic state space such as that of the restricted three-body problems (R3BP) poses a challenge to trajectory design. Unlike spacecraft motion in two-body dynamics, a three-body system causes the trajectory to revert into a chaotic state that heavily relies on its initial conditions. The unpredictable nature of a restricted three-body system poses many difficulties in trajectory design, yet its implications are proven extremely useful in space exploration. Due to its leveraging of the natural system dynamics, this type of trajectory typically requires a large transfer time and a relatively small $\Delta v$. By exploiting the dynamics near liberation points and weak stability boundaries, we can derive useful information to predict optimized, low-energy space travel. 

Weak stability boundary transfers (WSBTs) are a particular type of trajectory that leverages the inherently chaotic nature of the dynamics to the spacecraft's advantage to transit into or out of the vicinity of a primary body in the R3BP. 
In recent decades, researchers have made many attempts to design families of ballistic trajectories in R3BP with WSBTs starting with Belbruno in 1987 \cite{belbruno1987lunar}. The concept was first exemplified in JAXA's Hiten mission through a ballistic capture in a lunar orbit designed by Belbruno and Miller \cite{belbruno1993sun}. This capture relied only on the natural gravitational forces of the Earth, Moon, and Sun, making it an interesting alternative for lunar trajectories. Another type of ballistic capture discovered by Belbruno in 1987 was used in ESA’s SMART-1 mission \cite{schoenmaekers2001smart}. 
Since then, researchers have elaborated on WSBTs in multiple dynamical systems and discussed the possibility of permanent capture \cite{belbruno2004existence}. In 2006, Garcia \cite{garcia2007note} made attempts to clarify the algorithmic definition in a planar restricted three-body problem. 
Topputo and Belbruno \cite{topputo2009computation,topputo2015earth} expanded on this by considering the dynamical system between the Sun and Jupiter in a circular restricted three-body problem \cite{topputo2009computation} and Earth and Mars in an elliptical three-body system \cite{topputo2015earth}. From a mission design-driven perspective Hyeraci and Topputo \cite{Hyeraci2010} presented a systematic scheme for designing WSBT captures upon arrival to a planet at the end of interplanetary transfers. 
%
%
%
%

From the context of operating space assets in these chaotic domains, it is critical to identify not only a nominal WSBT along which a spacecraft may be flying, but also the state space surrounding this nominal. In fact, during an actual flight, the perfect state of the spacecraft is unattainable, and one must content with some state estimate anyway. 
There are two possible scenarios where the dynamics in the neighborhood of an object flying along a WSBT may be of interest. The first scenario is for navigation of this flying spacecraft, where it is desirable to know how long it would take for a neighbor state to diverge from the nominal under the order of magnitude of error in the state estimate. This would quantify a time scale within which additional measurements should be taken to update the state estimate, and potentially execute trajectory correction maneuvers should the spacecraft be on a diverging path. 
The second scenario considers the detection of non-cooperative targets. As an example, consider a debris drifting from the cislunar region to the translunar region on a WSBT. It is desirable to discern its path as early as possible, but orbit determination assets may be a limited resource. By providing a time-scale to have to wait before neighborhood trajectories become discernible from a nominal, we can provide insight that may be leveraged from a sensor tasking perspective. 

A number of techniques drawing from dynamical systems theory exist for analyzing the flow of a region in state space. 
One numerical quantity that forms the foundation of a multitude of other metrics is the Cauchy Green Tensor (CGT). The CGT is formed from the state transition matrix, and quantifies the naturally stretching directions of the flow. The CGT has been used in the context of trajectory design \cite{Oguri2020,Muralidharan2022} as well as station-keeping \cite{Guzzetti2017,Muralidharan2020}. Higher order methods, such as the use of state transition tensors, has also been studied \cite{Boone2023}. 
While eigenvectors of the CGT can offer directions that span the state space, it may also be desirable to obtain a scalar metric for the stretching nature. One such technique is the finite-time Lyapunov exponent (FTLE), which is a function of the largest eigenvalue of the CGT normalized by the propagation time. 
Kikuchi et al \cite{Kikuchi2019} studied the stability of coupled orbit-attitude dynamics through FTLE maps; Canales et al \cite{Canales2021} utilizes contours of FTLE for transfer design in a multi-Moon tour application with transits through the neck region of Jacobi contours. 
Finally, Lagrange coherent structures (LCS), typically constructed by looking at ridges in FTLE maps, can be employed for classifying regions in the state space \cite{haller2002lagrangian}. 
Recent work has discussed the implications of LCS in space mission design, specifically in R3BP trajectories \cite{shang2017trajectory}. Raffa et al \cite{raffa2023finding} and Tyler and Wittig \cite{tyler2022three} explored the use of Lagrangian descriptors and their success in locating bounded regions in non-autonomous dynamics models such as binary asteroid environments \cite{raffa2023finding} and Sun-Mars ER3BP \cite{tyler2022three}. By tracking time-dependent analogs of a non-autonomous system and comparing adjacent trajectories, propellant-efficient trajectory patterns and families of weak stability boundary transfers (WSBTs) can be extracted \cite{gawlik2009lagrangian}. 


In this work, we formulate a procedure for identifying WSBTs, and analyzing their sensitivity along their corresponding trajectory with varying parameters. With the aforementioned two scenarios in mind, we construct FTLE grids around selected locations along a nominal WSBT. This grid is constructed by leveraging the eigenvector of the 6D CGT. Specifically, we choose the most unstable eigenvectors to consider the navigation application, as the ``worst-case'' in this scenario would correspond to the fastest divergence of a neighbor state from the nominal. In contrast, the most stable eigenvectors are used for the detection application, since the ``worst-case'' here would correspond to the slowest divergence of a neighbor state, thereby hindering subsequent, more precise detection. 
Along with FTLE, we also introduce the notion of separation time. This quantifies, for a given neighbor state, the time it takes to get outside of some 6D ball around the nominal path. By setting this 6D ball to correspond to state estimate accuracy in position and velocity that can be obtained through orbit determination, this metric allows us to obtain the time scale that drives the decision for subsequent measurement collection in both navigation and detection scenarios. 

The remaining of this paper is organized as follows: firstly, an overview of the dynamical systems theory relevant to this work is introduced. Namely, these are the equations of motion, the FTLE, and the WSBT. This is followed by a description of the method employed in this paper. Specifically, emphasis is placed on the grid construction based on the CGT eigenvector, and the notion of the separation time. 
We then present numerical results based on WSBTs in the Earth-Moon ER3BP.
Finally, a conclusion is provided summarizing the work. 
%
%

\section{Dynamical System}
To explore the dynamics of an elliptical restricted three-body problem (ER3BP), the equations of motion are first established. To this end, we first introduce the circular restricted three-body problem (CR3BP) and extend this to the ER3BP. The elliptical assumption for the primary bodies in the ER3BP is what renders the dynamics to be non-autonomous. 
We then introduce the FTLE, which provides a scalar quantity for assessing the stretching nature of the dynamics. Finally, an overview of WSBTs and the procedure for generating a simple data set is given.

\subsection{Equations of Motion}
To consider a trajectory pattern in a non-autonomous system, it is important to first consider the dynamics in the CR3BP. The CR3BP considers the motion of an infinitesimal mass $m_3$ in the presence of the gravitational field of two large masses, $m_1 = -\mu$ and $m_2 = 1-\mu$, rotating about their center of mass \cite{gawlik2009lagrangian}. The motion of $m_3$ is constrained to an energy surface that links to the initial energy of the system. The units are normalized, and the position of all objects is assigned based on a coordinate frame that rotates counterclockwise with a unit angular frequency. Since it is a circular orbit, the frame rotates uniformly with a constant effective potential. The frame consists of $m_1$ and $m_2$ along the $x$-axis with the center of mass at the origin. There are five Lagrangian equilibrium points (L1-L5) that correspond to the critical points of the effective potential. Three of these equilibrium points lie co-linearly with $m_1$ and $m_2$ and the other two lie at the vertices of an equilateral triangle whose base connects $m_1$ and $m_2$.  


Let the state of the spacecraft $\boldsymbol{x} \in \mathbb{R}^6$ contain its Cartesian position and velocity. 
The CR3BP equations of motion are given by
\begin{equation}
    \begin{aligned}
        \ddot{x}-2 \dot{y} & = \dfrac{\partial U}{\partial x} \\
        \ddot{y}+2 \dot{x} & = \dfrac{\partial U}{\partial y} \\
        \ddot{z} & = \dfrac{\partial U}{\partial z}
    \end{aligned}
    \label{eq:eom_cr3bp}
\end{equation}
where $U$ is the effective potential function, given by
\begin{equation}
    U = \frac{1}{2}(x^{2}+y^{2})+\frac{1-\mu}{r_1}+\frac{\mu}{r_2}
    \label{eq:potential_cr3bp}
\end{equation}
with $\mu$ representing the mass parameter and
\begin{equation}
    \begin{aligned}
        r_1^{2}&=(x+\mu)^{2}+y^{2}+z^{2}\\
        r_2^{2}&=(x-1+\mu)^{2}+y^{2}+z^{2}
    \end{aligned}
    \label{radii}
\end{equation}
where $r_i$ $(i=1,2)$ are the distances from $m_3$ to $m_1$ and $m_2$ respectively.
In ER3BP, a non-autonomous extension of the circular R3BP, the dynamics differ due to the primaries' eccentricity, causing a variance in effective potential with respect to time. 
To ensure the position of $m_1$ and $m_2$ remains constant, we may consider a synodic, pulsating coordinate frame that rotates non-uniformly. In addition, the motion of 
$m_3$ is no longer constricted to the formally three-dimensional energy surface and must now be considered in a fourth dimension. 
For further discussion on the ER3BP dynamics, the reader is directed to the literature \cite{Broucke1969,Gomez1986,Hyeraci2010,Ferrari2018}. Let $\omega$ be the pseudo-potential of the ER3BP, 
\begin{equation}
\label{eq:potential_er3bp}
    \omega = \frac{1}{1+e \cos(f)}\Omega
\end{equation}
\begin{equation}
    \Omega = \frac{1}{2}\left( x^{2} + y^{2} \right) + \frac{1-\mu}{r_{1}} + \frac{\mu}{r_{2}} + \frac{1}{2}\mu(1-\mu) + \frac{1}{2}e \cos(f)z^{2}
\end{equation}
then the ER3BP equations of motion are given by
\begin{equation}
    \begin{aligned}
        x^{\prime \prime}-2 y^{\prime} & = \dfrac{\partial \omega }{\partial x} \\
        y^{\prime \prime}+2 x^{\prime} & = \dfrac{\partial \omega }{\partial y} \\
        z^{\prime \prime} & = \dfrac{\partial \omega }{\partial z} 
    \end{aligned}
    \label{eq:eom_er3bp}
\end{equation}
where $(\cdot)^{\prime}$ indicate derivatives with respect to the true-anomaly $f$ of the primaries.

\subsection{Cauchy-Green Tensor and Finite-Time Lyapunov Exponent}
The FTLE is the time-varying analog to invariant manifolds for non-autonomous dynamics. 
It is a scalar metric that quantifies the degree to which the flow of the dynamics stretches integrated over a given time span. 
Let the flow map be defined as $\phi^{t}_{t_0}(x)$ which provides the given state at time $t$ evolved from an initial point $x$ at time $t_0$. The state of this map heavily relies on its evolution from the initial conditions of the system. The Jacobian matrix ${\text{d}\phi^{t}_{t_0}(x)}/{\text{d}x_0}$ represents the state transition matrix (STM). This can be obtained by integrating the ER3BP equations of motions over a specified time, typically coordinating with the system's period. 
Let the Cauchy-Green strain tensor (CGT) $\Delta$ be defined as
\begin{equation}
    \Delta = \left[\frac{\text{d}\phi^{t}_{t_0}(\boldsymbol{x})}{\text{d} \boldsymbol{x}_0}\right]^{T}\left[
        \frac{\text{d}\phi^{t}_{t_0}(\boldsymbol{x})}{\text{d} \boldsymbol{x}_0}\right]
    \label{CG_strain_er3bp}
\end{equation}
The FTLE $\sigma(\phi^{t}_{t_0}; \boldsymbol{x})$ for a given propagation time $T$ and state $\boldsymbol{x}$ is given by
\begin{equation}
    \sigma(\phi^{t}_{t_0}; \boldsymbol{x}) = \dfrac{1}{|T|} \log \sqrt{\lambda_{\max}(\Delta(\boldsymbol{x}))}
\end{equation}
where $\lambda_{\max}$ is the largest eigenvalue of $\Delta$. 
The STM ${\text{d}\phi^{t}_{t_0}(\boldsymbol{x})} / {\text{d} \boldsymbol{x}_0}$ is propagated using the Jacobian of the dynamics $\boldsymbol{A}(\boldsymbol{x}(t))$, resulting in the initial value problem
\begin{equation}
    \begin{aligned}
        \dfrac{\mathrm{d}}{\mathrm{d}t} \left[ \frac{\text{d}\phi^{t}_{t_0}(\boldsymbol{x})}{\text{d} \boldsymbol{x}_0} \right]
        &= \boldsymbol{A}(\boldsymbol{x}(t)) \frac{\text{d}\phi^{t}_{t_0}(\boldsymbol{x})}{\text{d} \boldsymbol{x}_0}
        \\
        \frac{\text{d}\phi^{t_0}_{t_0}(\boldsymbol{x})}{\text{d} \boldsymbol{x}_0} &= \boldsymbol{I}_6
    \end{aligned}
\end{equation}
where the Jacobian is given by
\begin{equation}
    \boldsymbol{A}(\boldsymbol{x}(t)) = \left[
    \begin{array}{c|c}
    \boldsymbol{0}_{3,3} & \boldsymbol{I}_{3,3} \\
    \hline
    \omega_{\boldsymbol{x} \boldsymbol{x}} & \boldsymbol{W}
    \end{array}
    \right]
    \,,\quad 
    \boldsymbol{W} = \begin{bmatrix}
        0  &  2  &  0 \\
        -2 &  0  &  0 \\
        0  &  0  &  0 
     \end{bmatrix}
\end{equation}
where $\omega_{\boldsymbol{x} \boldsymbol{x}} \in \mathbb{R}^{3 \times 3}$ is the second-order partials of the pseudo-potential $\omega$. 
We note that since the dynamics is defined with respect to true anomaly derivatives, the notion of time must be considered as elapsed true anomaly angular separation. However, since the system eccentricity is typically small, we employ this quantity as a proxy to the actual, physical time for the analysis in subsequent Sections. 

\subsection{Weak Stability Boundary Transfers}
WSBTs reside in regions that are at the boundary of the gravitational influence of two primaries. Borrowing from the two-body dynamics lexicon (with a slight abuse of terminology), this corresponds to the vicinity of the sphere of influence of the smaller primary. 
WSBTs are constructed following the approach proposed by Belbruno et al \cite{BELBRUNO20081330}. Assuming planar motion, a grid of initial states are constructed by assigning a pair of radius and velocity, where the velocity is assumed to be purely tangential and counter-clockwise. 
The velocity magnitude is obtained by prescribing an eccentricity $0 \leq e < 1$ and assuming the spacecraft is at its periapsis on an osculating Keplerian orbit. 


\section{Method for Separation Time Analysis}
This Section discusses the method explored in this work to analyze the sensitivity along WSBTs. 
First, a description on the procedure to generate a grid around a given state along the nominal WSBT. 
This is followed by an introduction on the concept of the separation time. 

\subsection{Grid Construction}
The grid is constructed using the eigenvector of the CGT $\Delta\in\mathbb{R}^{6\times6}$ for a given nominal state $\bar{\boldsymbol{x}}\in\mathbb{R}^6$ on the WSBT with some prescribed propagation time $T$. Then, defining the corresponding eigenvector respective to the largest or smallest eigenvalue as $\boldsymbol{\nu}_{\max/\min}\in\mathbb{R}^6$, the grid around the nominal state is constructed by perturbing the position and velocity along the position space of the eigenvector, $\boldsymbol{\nu}_{r,\,\max/\min}\triangleq \boldsymbol{\nu}_{\max/\min}\left[0:3\right]\in\mathbb{R}^3$, and the velocity space of the eigenvector, $\boldsymbol{\nu}_{v,\,\max/\min}\triangleq \boldsymbol{\nu}_{\max/\min}\left[3:6\right]\in\mathbb{R}^3$, respectively. Here, $\boldsymbol{\nu}_{\max/\min}\left[0:3\right]$ and $\boldsymbol{\nu}_{\max/\min}\left[3:6\right]$ are the first and last three elements of $\boldsymbol{\nu}_{\max/\min}\in\mathbb{R}^6$. 
These vectors are perturbed along the two directions multiplied by some scalar $\pm \epsilon_r \delta r_{\mathrm{OD}}$ and $\pm \epsilon_v \delta v_{\mathrm{OD}}$, where $\epsilon_{r,v} \in (0,1]$. 
This process of computing the grid is summarized as follows:
\begin{enumerate}
    \item Choose nominal state $\bar{\boldsymbol{x}}$ at some time $t_0$ along the nominal WSBT
    \item Compute $\Delta$ from $\bar{\boldsymbol{x}}$ for some $T$
    \item Compute eigenvector corresponding to largest or smallest eigenvalue $\boldsymbol{\nu}_{\max/\min}$. The position and velocity directions defined as 
    \begin{equation}
    \begin{aligned}
        \boldsymbol{\nu}_{r,\,\max/\min} &=
        \boldsymbol{\nu}_{\max/\min}\left[0:3\right]
        \\
        \boldsymbol{\nu}_{v,\,\max/\min} &=
        \boldsymbol{\nu}_{\max/\min}\left[3:6\right]
    \end{aligned}
\end{equation}
    \item For $\epsilon_r \in (0,1]$ and for $\epsilon_v \in (0,1]$, construct neighbor state $\boldsymbol{x}$ via
    \begin{equation}
        \boldsymbol{x} = \bar{\boldsymbol{x}} \pm
        \begin{bmatrix}
            \epsilon_r \delta r_{\mathrm{OD}}
            \boldsymbol{\nu}_{r,\,\max/\min}
            \\
            \epsilon_v \delta v_{\mathrm{OD}}
            \boldsymbol{\nu}_{v,\,\max/\min}
        \end{bmatrix}
    \end{equation}
\end{enumerate}
The choice on whether to use the largest or smallest eigenvectors depends on the application of interest; if the aim is to determine the worst-case, fastest divergence for navigation applications, then the largest eigenvectors, corresponding to the largest stretching directions, should be used. Meanwhile, if the aim is to determine the worst-case, slowest separation time that hinders the performance of detection of non-cooperative targets, then the smallest eigenvectors, corresponding to the smallest stretching directions, would be appropriate.

\subsection{Separation Time}
\begin{figure}
    \centering
    \includegraphics[width=0.92\linewidth]{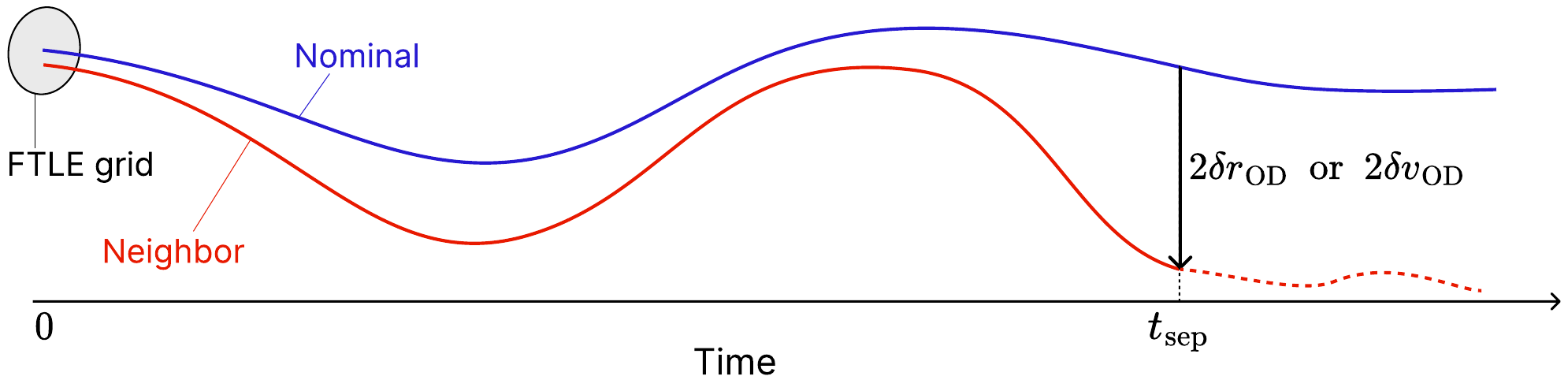}
    \caption{Illustration of separation time. The blue nominal trajectory originates from some point (usually the center) of the grid; the red neighbor trajectory stems from one of the gridded state, and its evolution in time is tracked. When the Euclidean distance of the position or velocity vector exceeds a threshold, the time stamp is recorded as the separation time for this neighbor trajectory.}
    \label{fig:sep_time_concept}
\end{figure}

To obtain a scalar metric for studying a nominal state's sensitivity, the separation time $t_{\mathrm{sep}}$ is introduced. Let $\bar{\boldsymbol{r}}(t)$ and $\bar{\boldsymbol{v}}(t)$ be the nominal WSBT's position and velocity vectors at time $t$. Then, the separation time $t_{\mathrm{sep}}$ is defined as the minimum time required for either the position or velocity vector of a neighbor state, $\boldsymbol{r}(t)$ or $\boldsymbol{v}(t)$, to be separated by some threshold value. 
This can be expressed as 
\begin{equation}
    \begin{aligned}
        & t_{\mathrm{sep}} = 
        \\
        & \min \left(
        \underset{t}{\operatorname{argmin}} \left[
            \| \boldsymbol{r}(t_0+t) - \bar{\boldsymbol{r}}(t_0+t) \|_2 \geq 2\delta r_{\mathrm{OD}}
        \right]
        ,\,
        \underset{t}{\operatorname{argmin}} \left[
            \| \boldsymbol{v}(t_0+t) - \bar{\boldsymbol{v}}(t_0+t) \|_2 \geq 2\delta v_{\mathrm{OD}}
        \right]
    \right)
    \end{aligned}
    \label{eq:t_separation}
\end{equation}
Figure \ref{fig:sep_time_concept} shows the concept of the separation time for a neighbor trajectory. 
The separation threshold distance is chosen to be two times the orbit determination ball radii $
\delta r_{\mathrm{OD}}$ and $\delta v_{\mathrm{OD}}$; this accounts for the worst case scenario where the detected nominal falls at the edge of the uncertainty ball. 
Furthermore, if exactly the ball radius was used, the edge of the grid would likely immediately separate beyond the threshold, thus giving a $t_{\mathrm{sep}} \approx 0$, which does not give insight to the problem at hand.  

\subsection{Comparison between FTLE and Separation Time}
We generate a set of FTLE maps and separation time maps around the nominal trajectory based on the aforementioned grid, but there is a qualitative difference between them. 
The fundamental difference is that FTLE is independent of the nominal WSBT, while separation time explicitly quantifies a scalar metric concerning the nominal WSBT. This means that the FTLE characterizes the local flow behavior autonomously at each grid point, while separation time provides a measure of how neighboring dynamics deviate from the nominal WSBT. While both these contours offer valuable information to the numerical results, it is imperative to bear in mind these differences while interpreting the following analysis.

\section{Numerical Results}
The proposed analysis framework is employed to analyze WSBTs in the Earth-Moon ER3BP, with $\mu = 0.01215058426994$ and $e = 0.0549$. We first present the two selected WSBTs, which are used as study cases. 
Then, we present analysis using the FTLE based on the unstable eigenvectors of $\Delta_r$ and $\Delta_v$, assuming navigation applications for a cooperative spacecraft.
Thirdly, we present analysis using the FTLE based on the  stable eigenvectors, corresponding to applications involving detection of non-cooperative targets. 

\subsection{Selected Nominal Weak Stability Boundary Transfers}
\begin{figure}[h]
    \centering
    \includegraphics[width=0.90\linewidth]{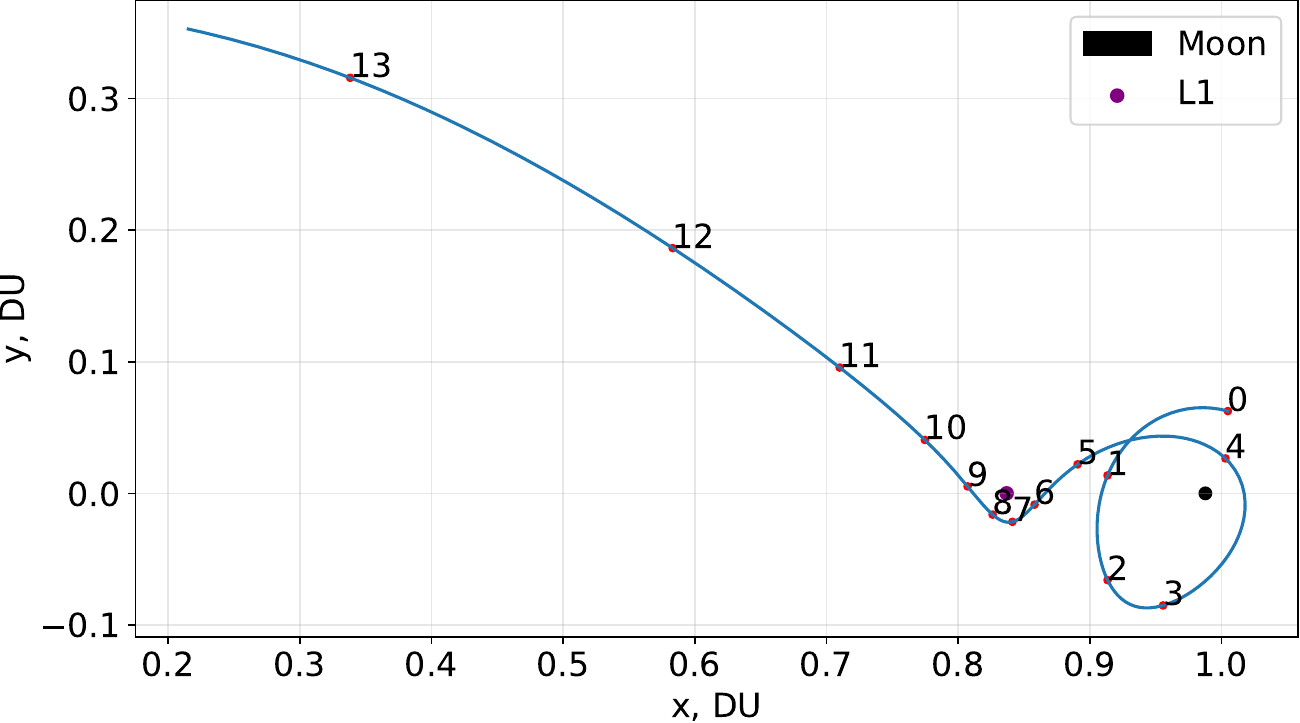}
    \caption{WSBT (a) shown in Earth-Moon rotatin-pulsating frame, with integer labels corresponding to time indices corresponding to locations where FTLE grid is constructed.}
    \label{fig:nom_wsbt_0}
\end{figure}

\begin{figure}[h]
    \centering
    \includegraphics[width=0.92\linewidth]{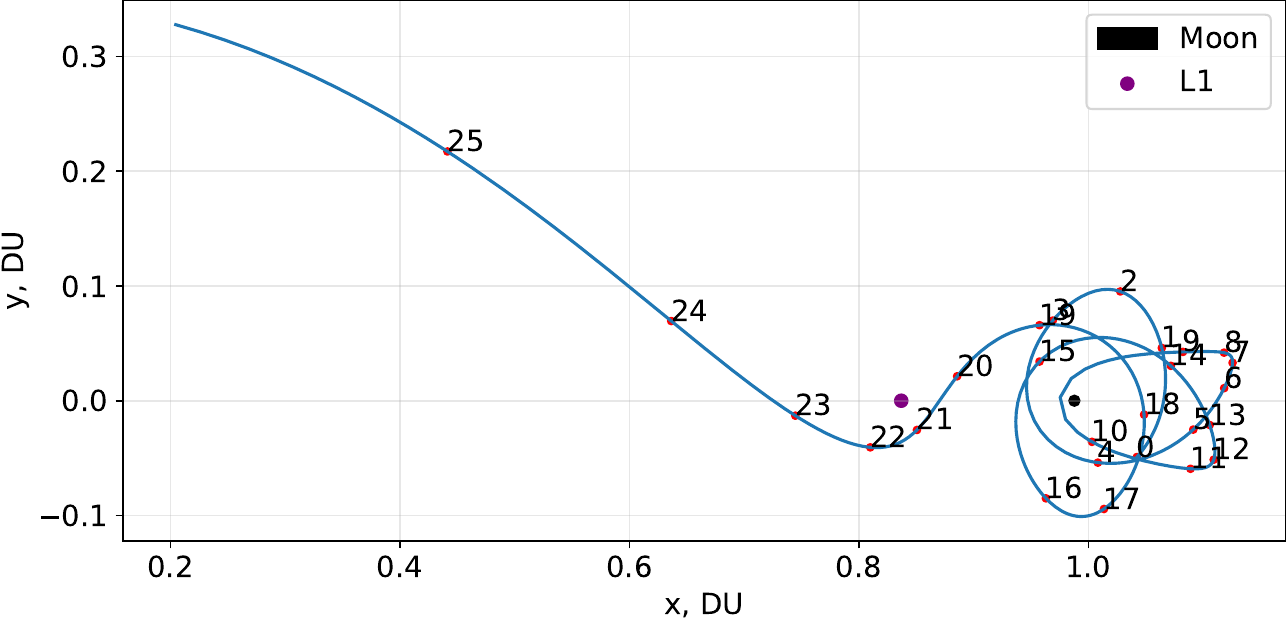}
    \caption{WSBT (b) shown in Earth-Moon rotating-pulsating frame, with integer labels corresponding to time indices corresponding to locations where FTLE grid is constructed.}
    \label{fig:nom_wsbt_18}
\end{figure}

To analyze the results pertaining to a nominal state's sensitivity, the nominal WSBTs shown in Figures \ref{fig:nom_wsbt_0} and \ref{fig:nom_wsbt_18} were chosen. The selected WSBTs offer two alternate routes around the L1 point. One of these trajectories exhibits a single revolution around the Moon before escape, while the other traverses multiple revolutions. 

\subsection{Navigation of Cooperative Spacecraft}
In the context of cooperative spacecraft, the optimal scenario involves maintaining a separation threshold distance within a radius of twice that of some 6D ball around the nominal path . To analyze its capacity to stay within its nominal's vicinity, we must consider when the nominal's state is most sensitive, specifically, the minimum separation time. To analyze the minimum separation time, we consider an unstable grid aligned with the largest eigenvector, denoted as $\boldsymbol{\nu}_{r,\,\max}$ and $\boldsymbol{\nu}_{v,\,\max}$. As this grid is perturbed along the direction of greatest stretching, it exhibits the most pronounced divergence from the nominal trajectory. Worth noting is that the results remain consistent across varying orbit determination ball radii, spanning from (1 km, 1 cm/s) to (100 km, 100 cm/s). As such, the ensuing analysis focuses on a ball radii of (10 km, 10 cm/s) to ensure coherence.

\begin{figure}[h]
    \centering
    \includegraphics[width=0.92\linewidth]{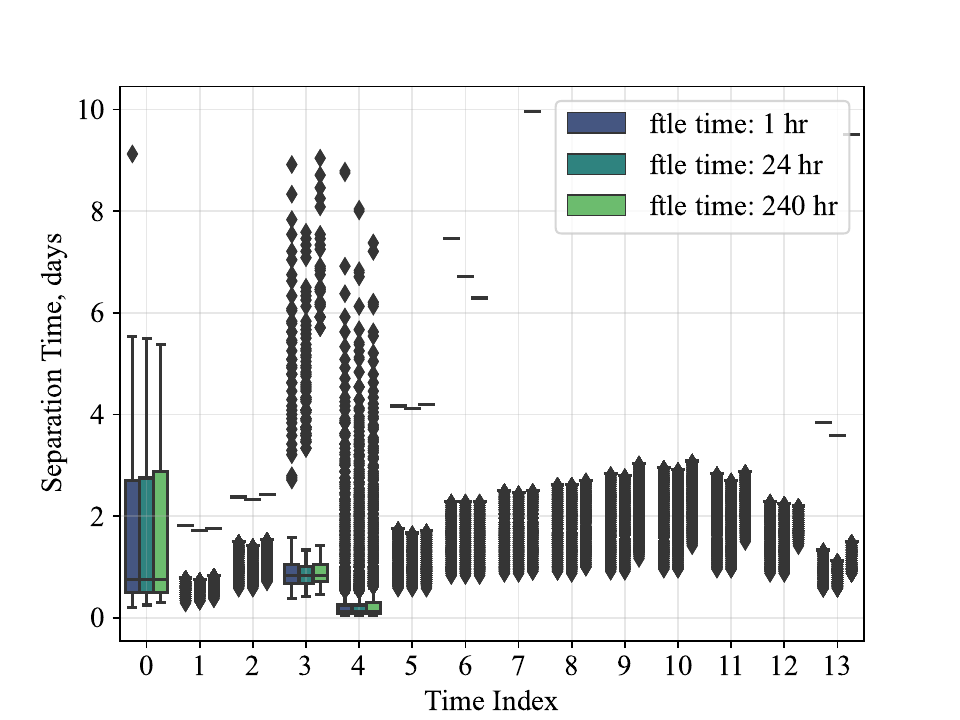}
    \caption{WSBT (a): Separation Time vs. Time Index (unstable grid,  OD-ball=(10 km, 10 cm/s))}
    \label{fig:00_ftle_box_unstable}
\end{figure}

\begin{figure}
    \centering

    \begin{subfigure}[b]{0.45\textwidth}
        \includegraphics[width=\textwidth]{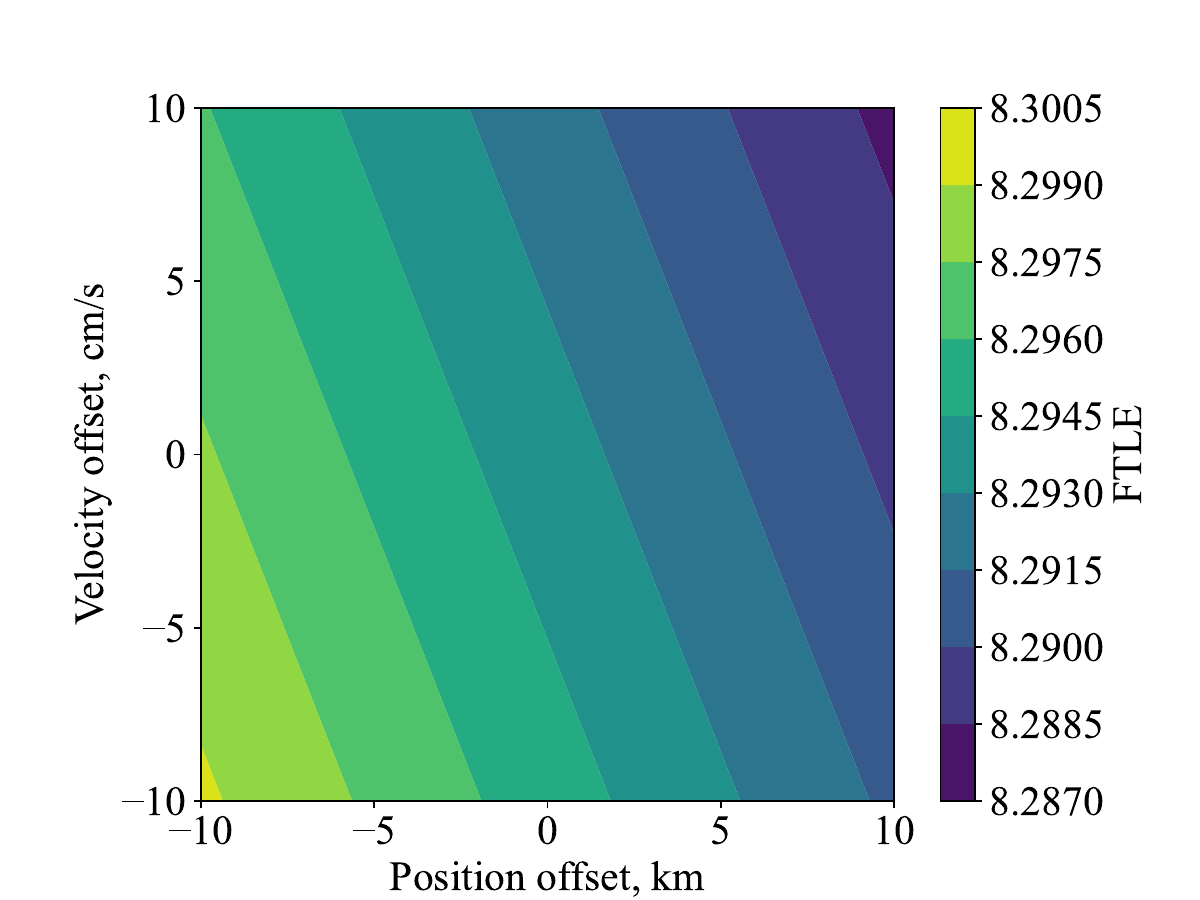}
        \caption{WSBT (a): FTLE at time index = 2, or $t=2.79$ day (T = 24.00 h, grid=unstable)}
        \label{fig:00_ftle_start}
    \end{subfigure}
    \hfill
    \begin{subfigure}[b]{0.45\textwidth}
        \includegraphics[width=\textwidth]{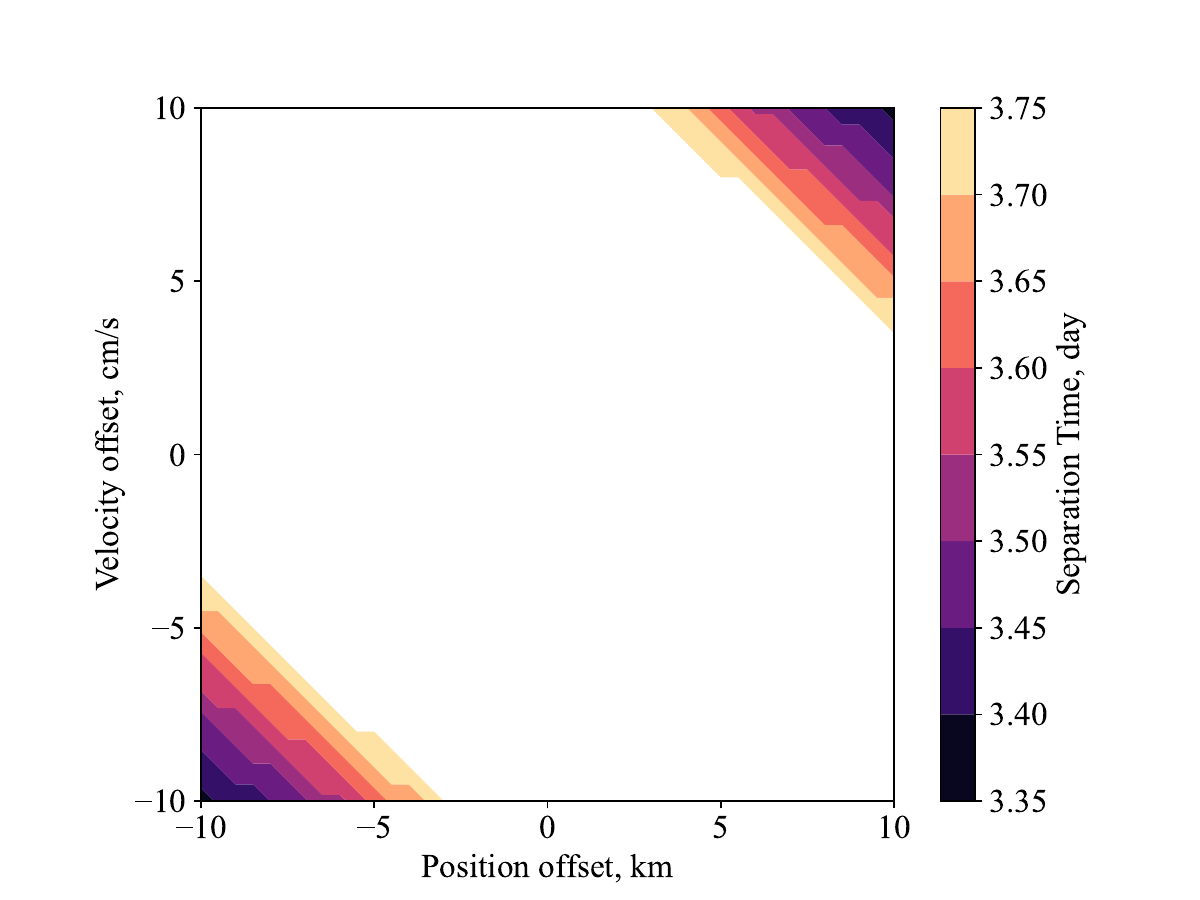}
        \caption{WSBT (a): Separation Time at time index = 2, or t=2.79 day ($T=24$ hours, unstable grid)}
        \label{fig:00_tsep_graph}
    \end{subfigure}

    \vspace{0.5cm} 

    \begin{subfigure}[b]{0.45\textwidth}
        \includegraphics[width=\textwidth]{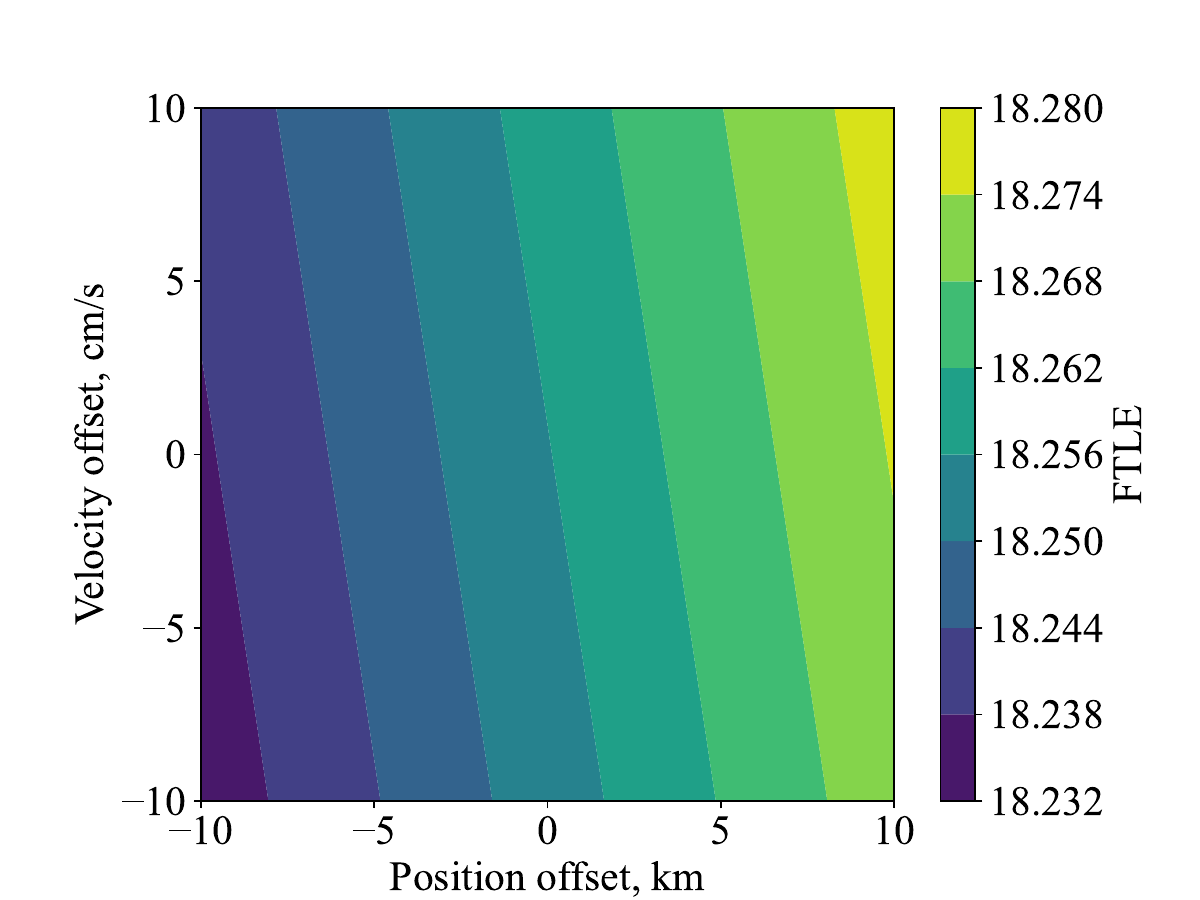}
        \caption{WSBT (a): FTLE at time index = 4, or t=5.57 day (T = 24.00 h, grid=unstable)}
        \label{fig:00_ftle_end}
    \end{subfigure}
    \hfill
    \begin{subfigure}[b]{0.45\textwidth}
        \includegraphics[width=\textwidth]{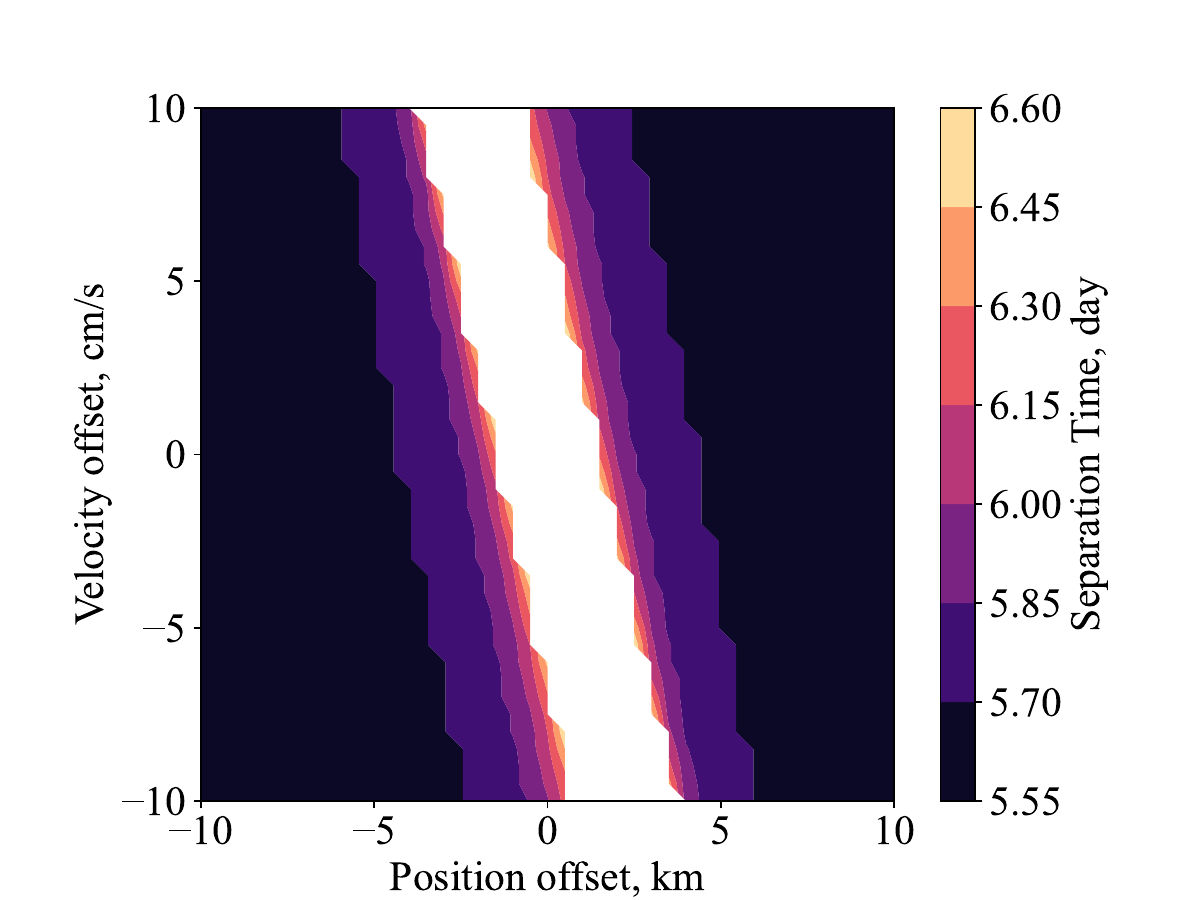}
        \caption{WSBT (a): Separation Time at time index = 4, or t=5.57 day ($T=24$ hours, unstable grid)}
        \label{fig:00_tsep_graph_end}
    \end{subfigure}

    \caption{Contour FTLE and Separation Time Plots for WSBT (a) with unstable grid}
    \label{fig:00_contour_unstable}
\end{figure}

Figure \ref{fig:00_ftle_box_unstable} exhibits a series of box and whisker plots for different propagation times along WSBT (a). At the nominal WSBT's perilune point, the minimum separation time nears zero. These results are expected due to the heightened influence of gravitation forces, thus causes significant perturbations and deviations in their trajectories. At the nominals apolune point (seen at time index = 2) and further along its trajectory, the minimum separation time increases relative to the perilune point. This progression signifies the diminishing influence of gravitational forces as the spacecraft traverses from its perilune point to regions farther along its orbital path.

\begin{figure}[h]
    \centering
    \includegraphics[width=0.92\linewidth]{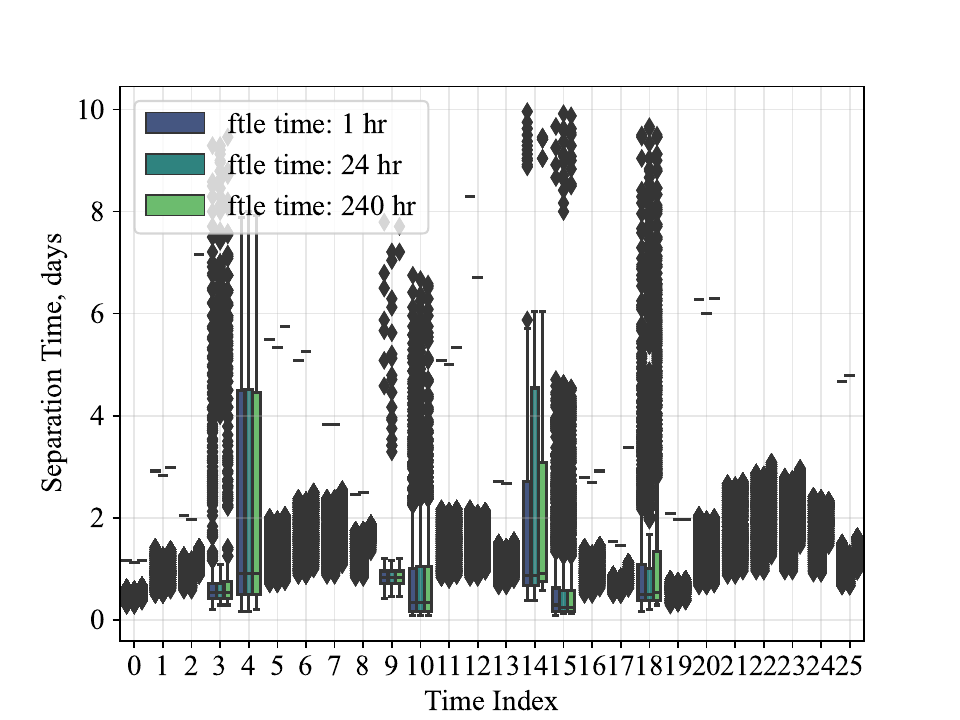}
    \caption{WSBT (b): Separation Time vs. Time Index (unstable grid,  OD-ball=(10 km, 10 cm/s))}
    \label{fig:18_ftle_box_unstable}
\end{figure}


\begin{figure}
    \centering

    \begin{subfigure}[b]{0.45\textwidth}
        \includegraphics[width=\textwidth]{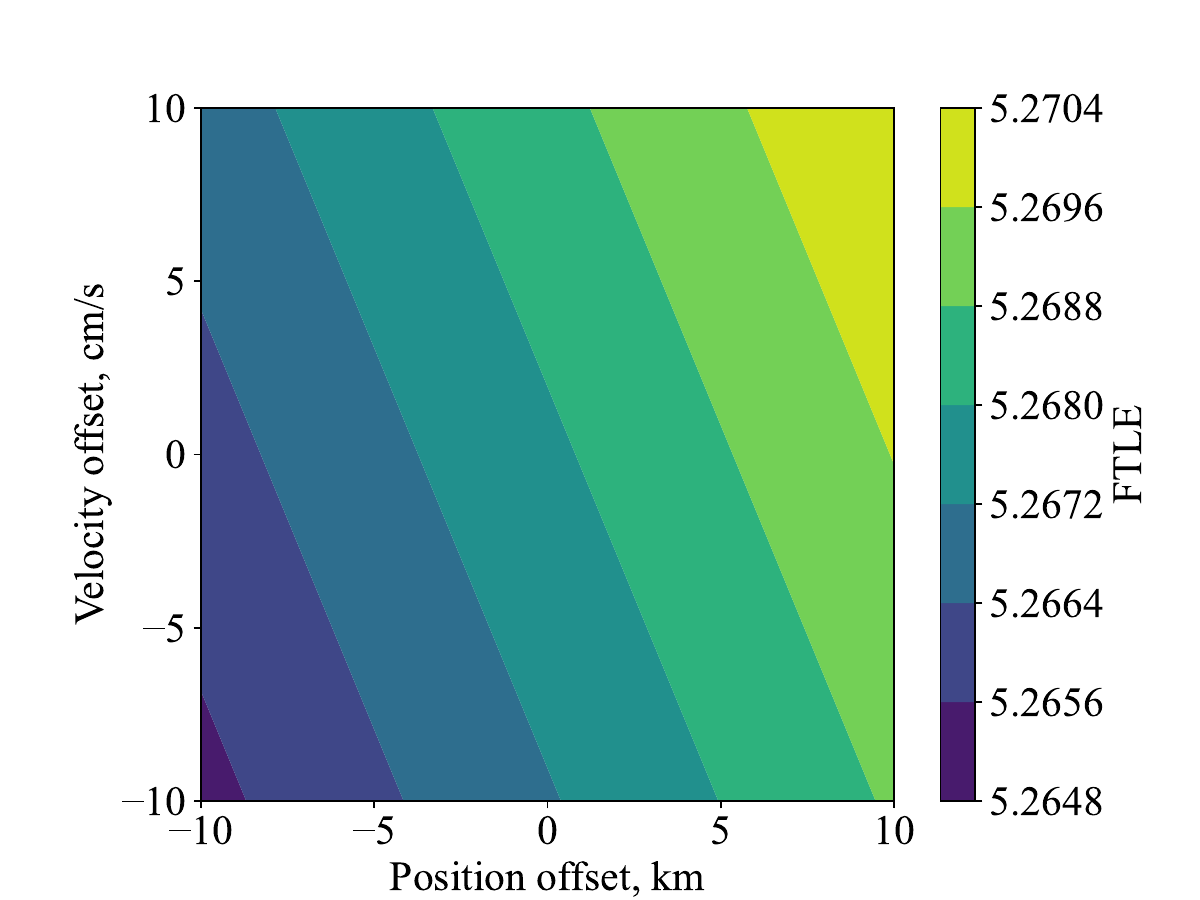}
        \caption{WSBT (b): FTLE at time index = 7, or t=9.75 day ($T = 24$ hours, grid=unstable)}
        \label{fig:18_ftle_start}
    \end{subfigure}
    \hfill
    \begin{subfigure}[b]{0.45\textwidth}
        \includegraphics[width=\textwidth]{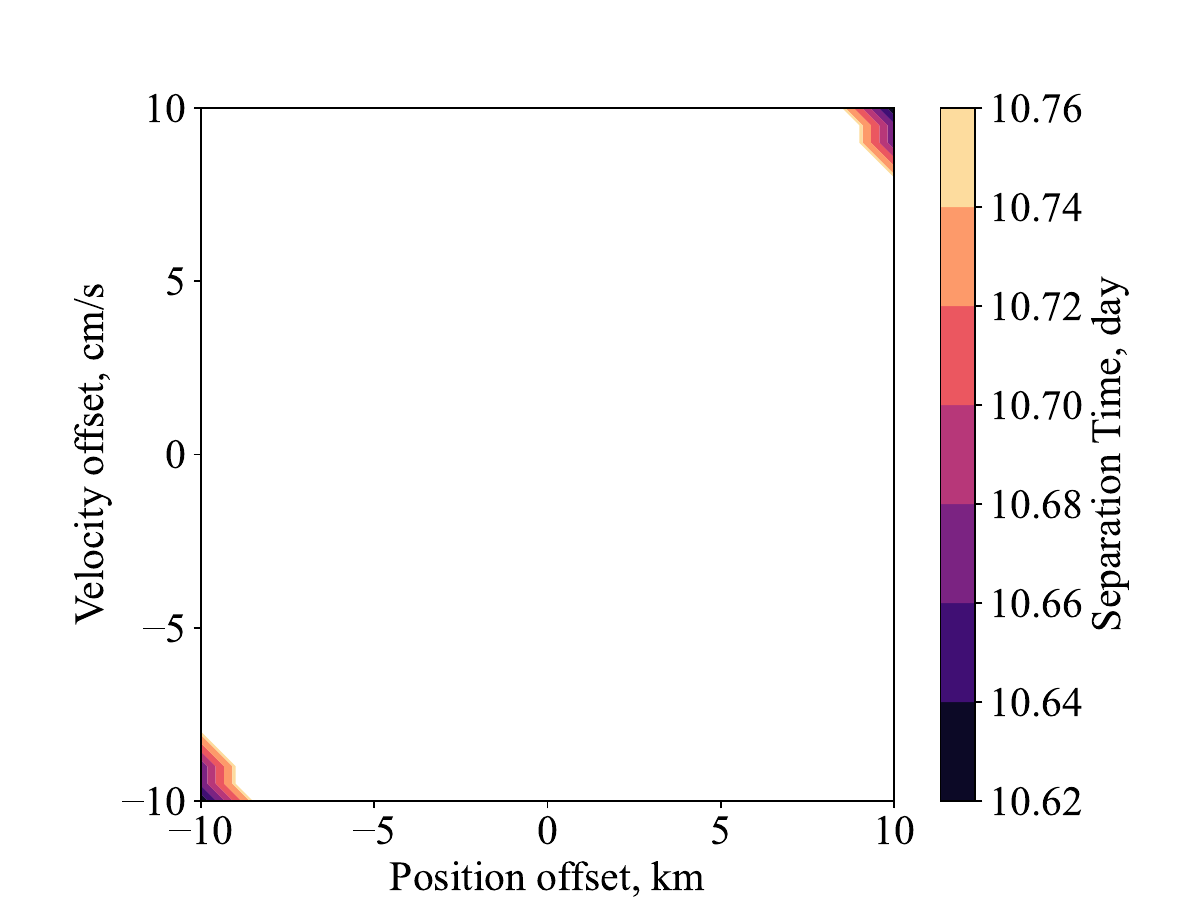}
        \caption{WSBT (b): Separation Time at time index = 7, or t=9.75 day ($T=24$ hours, unstable grid)}
        \label{fig:18_tsep_graph}
    \end{subfigure}

    \vspace{0.5cm} 

    \begin{subfigure}[b]{0.45\textwidth}
        \includegraphics[width=\textwidth]{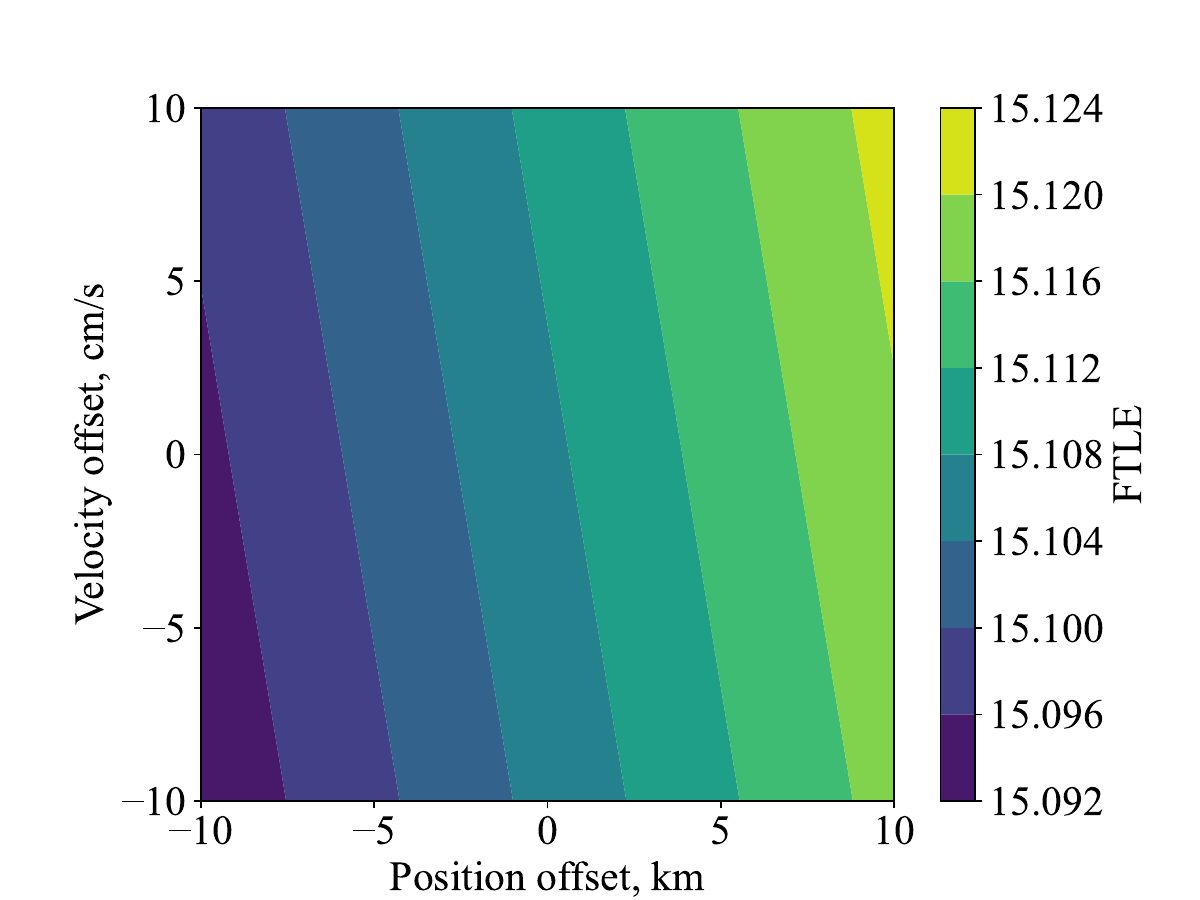}
        \caption{WSBT (b): FTLE at time index = 10, or t=13.93 day ($T=24$ hours, unstable grid)}
        \label{fig:00_ftle_end}
    \end{subfigure}
    \hfill
    \begin{subfigure}[b]{0.45\textwidth}
        \includegraphics[width=\textwidth]{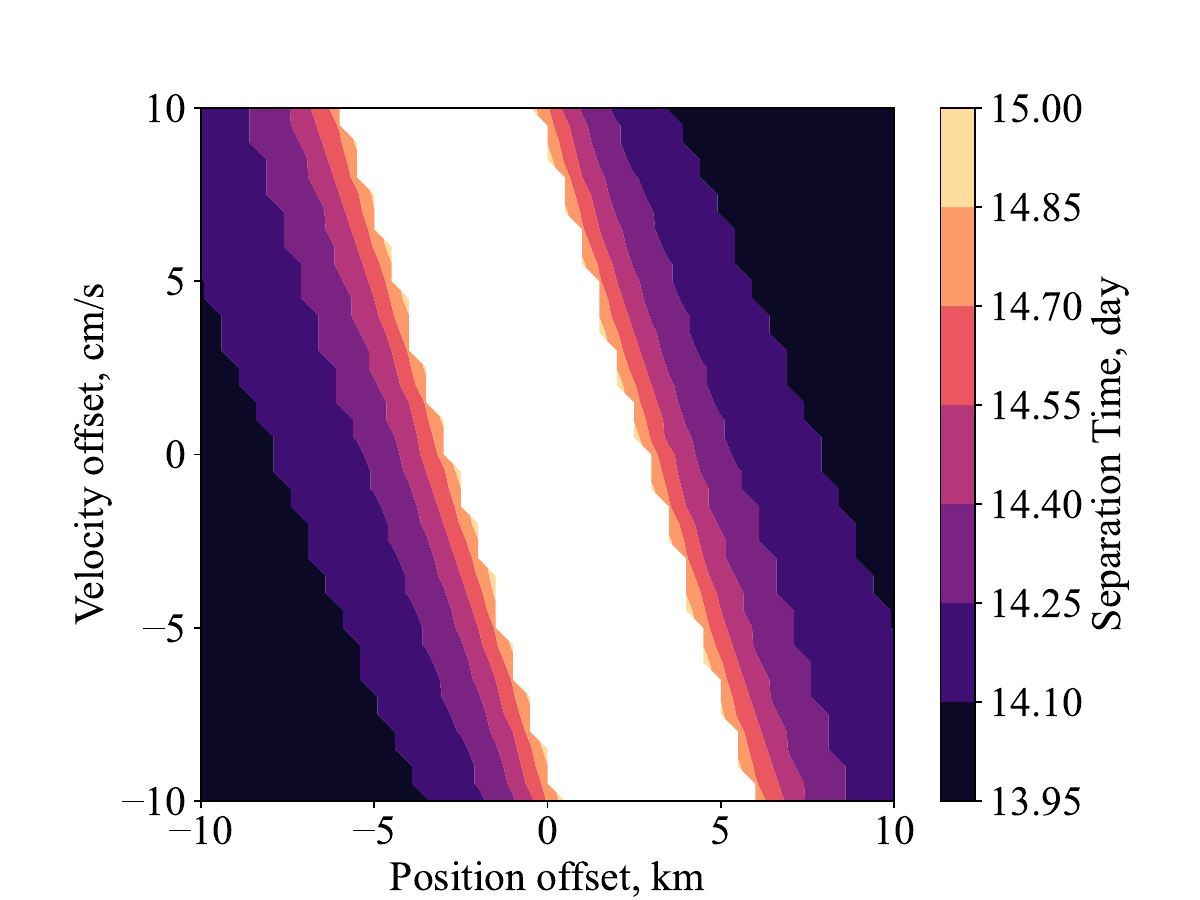}
        \caption{WSBT (b): Separation Time at time index = 10, or t=13.93 day ($T=24$ hours, unstable grid)}
        \label{fig:18_tsep_graph_end}
    \end{subfigure}

    \caption{Contour FTLE and Separation Time Plots for WSBT (b) with unstable grid}
    \label{fig:18_contour_unstable}
\end{figure}

In Figure \ref{fig:18_ftle_box_unstable}, there are relatively simlilar results to Figure \ref{fig:00_ftle_box_unstable} in which the minimum separation time decreases at its perilune points and increases points further from the Moon. Evidently, this leads us to conclude that the cooperative spacecraft is most susceptible to deviations from the nominal trajectory during its perilune points and with a certain proximity of the Moon. This analysis is further underscored in Figure \ref{fig:00_contour_unstable} and Figure \ref{fig:18_contour_unstable} where we examine the separation time contour nears the nominal WSBT's perilune and apolune points, accompanied by their corresponding FTLE grids. At a position distanced from the moon, shown in Figure \ref{fig:00_tsep_graph} and Figure \ref{fig:18_tsep_graph}, there are a higher proportion of points exhibiting no divergence, indicated by the white space. However, when the trajectory approaches its perilune state, shown in Figure \ref{fig:00_tsep_graph_end} and Figure \ref{fig:18_tsep_graph_end}, a substantial proportion of states indicate rapid divergence. Within the cooperative spacecraft context, the perilune state emerges as a pivotal juncture of heightened instability. Consequently, executing trajectory correction maneuvers during this phase poses notable challenges and underscores the need for precision in maneuver planning.

\subsection{Detection of Non-cooperative objects}
The stability, or lack thereof, of a nominal WSBT  holds significant implications for space domain awareness (SDA). In the context of detecting  non-cooperative objects in space, a highly unstable WSBT can yield enhanced awareness as the time of divergence becomes evident sooner. Opposed to the guidance and navigation of cooperative spacecraft, a relatively small time of separation is favored in SDA. When assessing our capability to track and identify space objects, we must examine points along a nominal trajectory characterized by longer separation times.

To analyze our ability to track and identify space objects, we must consider the points along a nominal with high separation times. To obtain these points, we consider a stable grid perturbed along a nominal state's smallest eigenvector, denoted as $\boldsymbol{\nu}_{r,\,\min}$ and $\boldsymbol{\nu}_{v,\,\min}$. As in previous tests, we consider a grid with prescribed $T=24$ hours and
an orbit determination ball radii of (10 km, 10 cm/s).

\begin{figure}[h]
    \centering
    \includegraphics[width=0.92\linewidth]{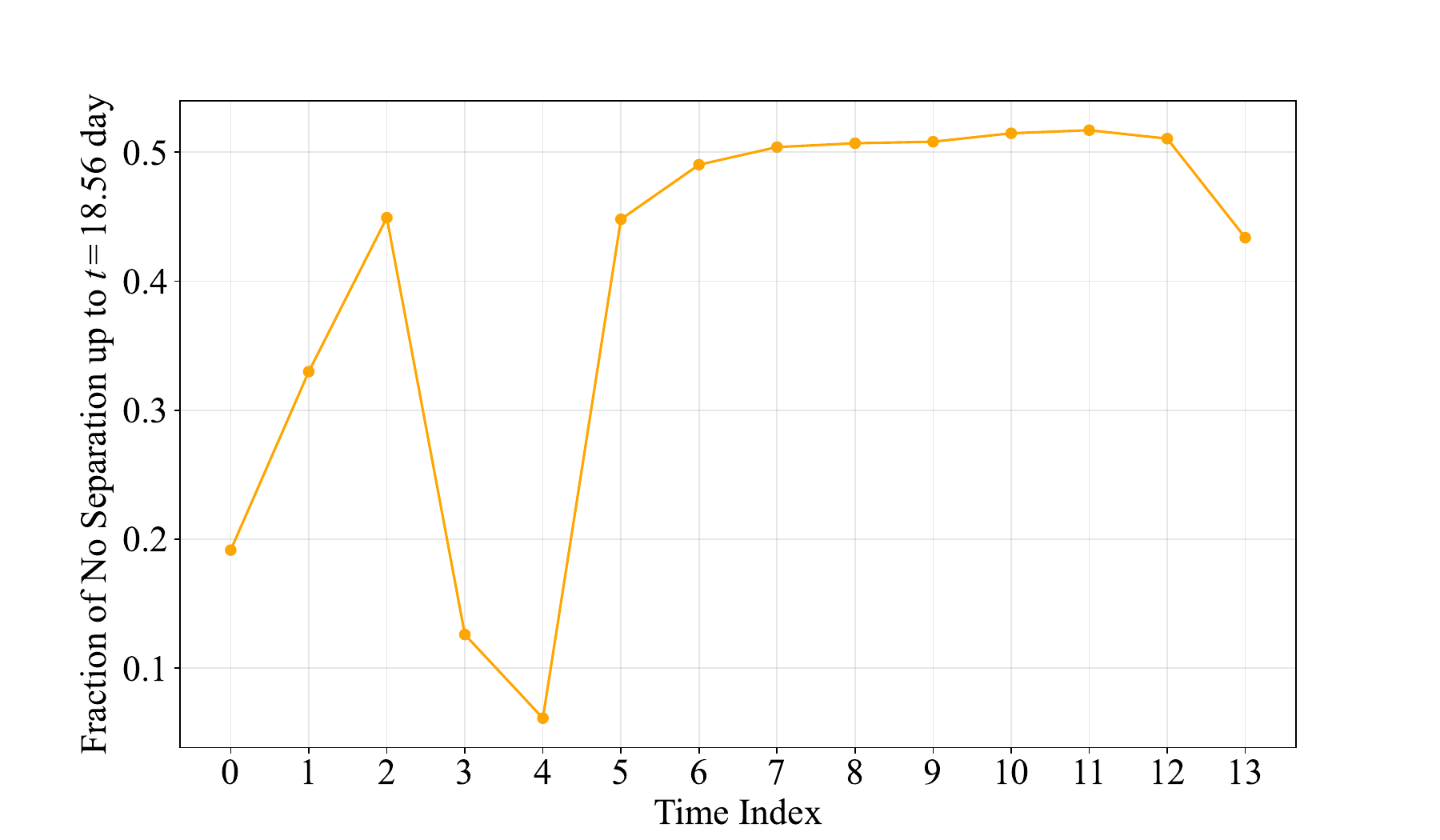}
    \caption{WSBT (a): Fraction of No Separation vs. Time Index ($T=24$ hours, stable grid, OD-ball=(10 km, 10 cm/s))}
    \label{fig:00_no_tsep_rate}
\end{figure}
\begin{figure}[h]
    \centering
    \includegraphics[width=0.92\linewidth]{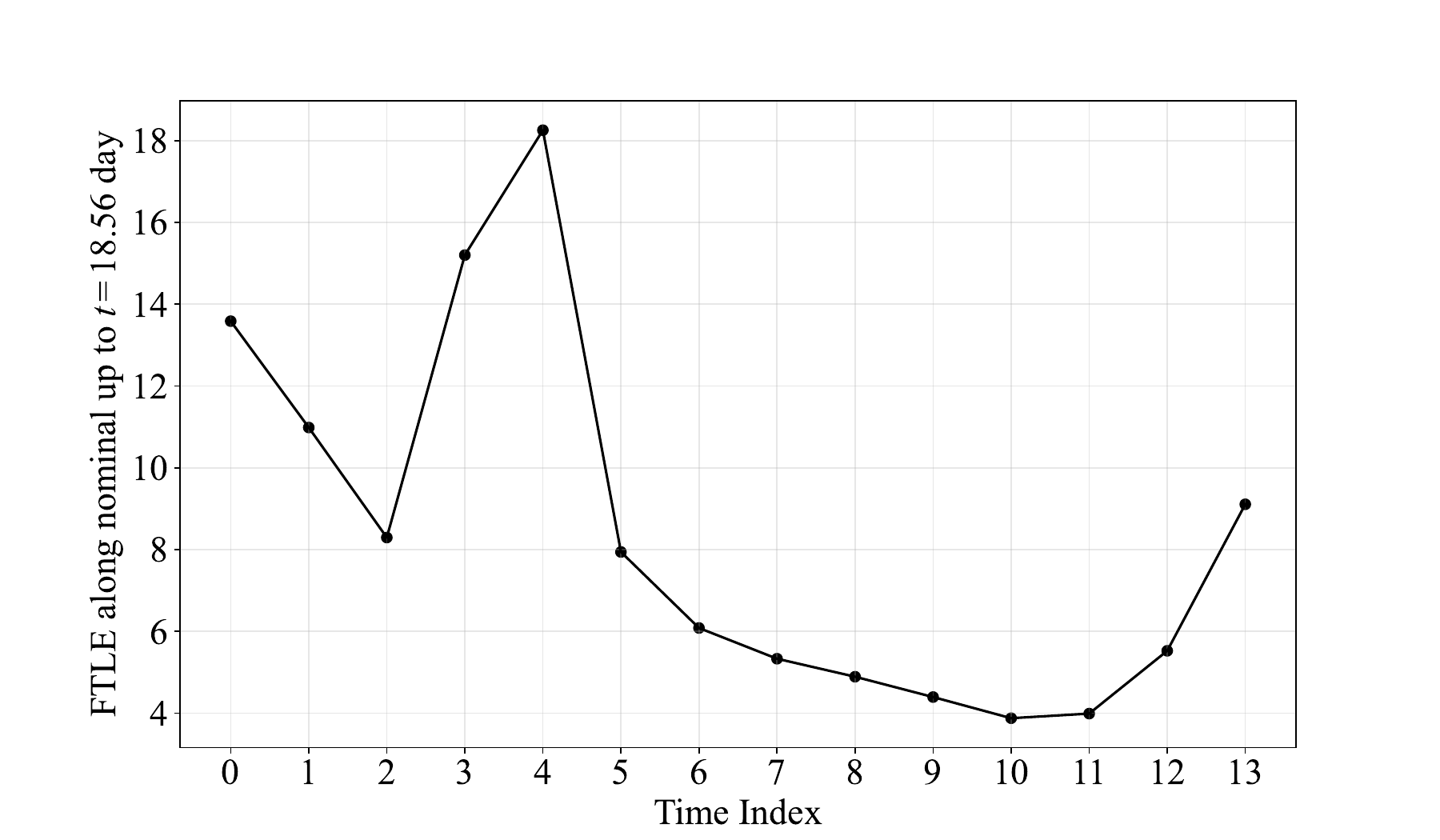}
    \caption{WSBT (a): FTLE along nominal ($T=24$ hours)}
    \label{fig:00_ftle}
\end{figure}

\begin{figure}
    \centering

    \begin{subfigure}[b]{0.45\textwidth}
        \includegraphics[width=\textwidth]{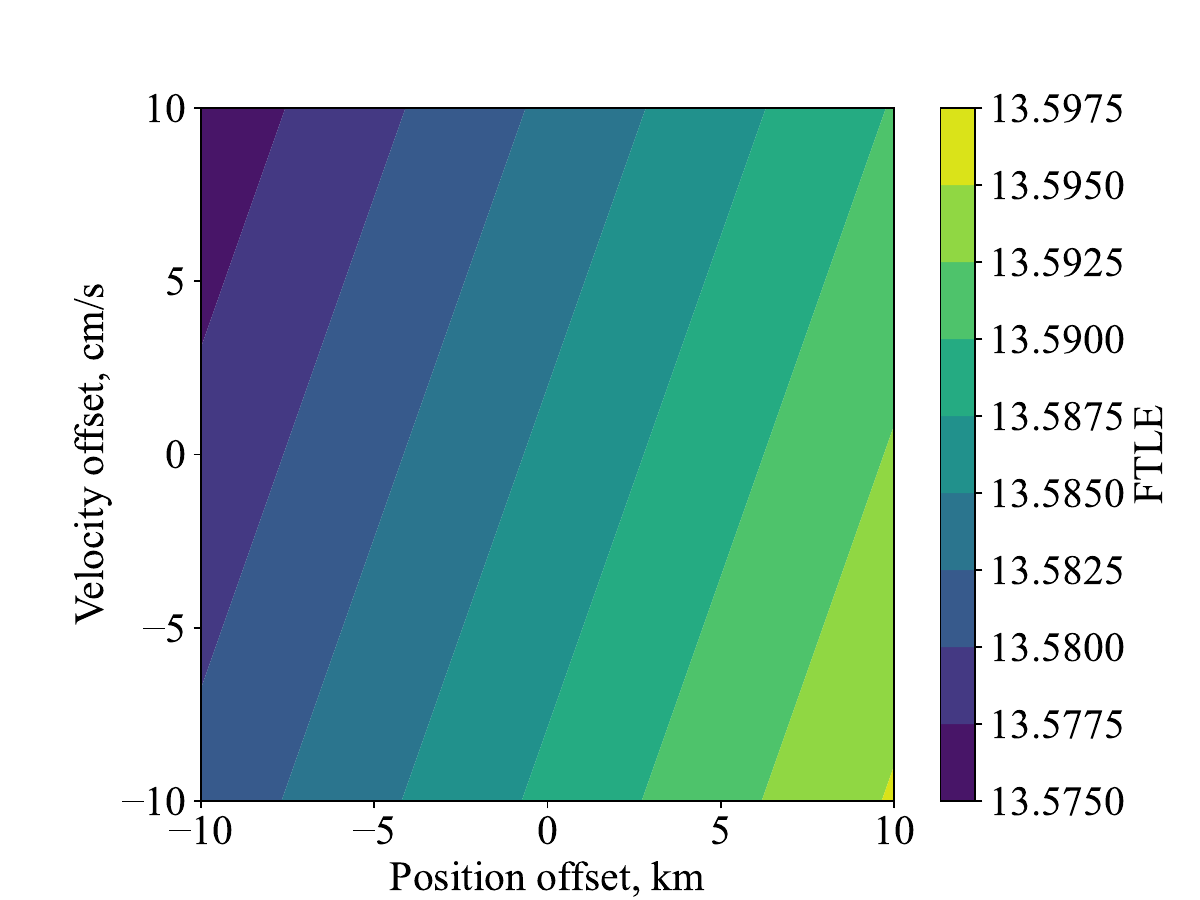}
        \caption{WSBT (a): FTLE at time index = 0, or t=0.00 day ($T=24$ hours, stable grid)}
        \label{fig:00_ftle_start_stable}
    \end{subfigure}
    \hfill
    \begin{subfigure}[b]{0.45\textwidth}
        \includegraphics[width=\textwidth]{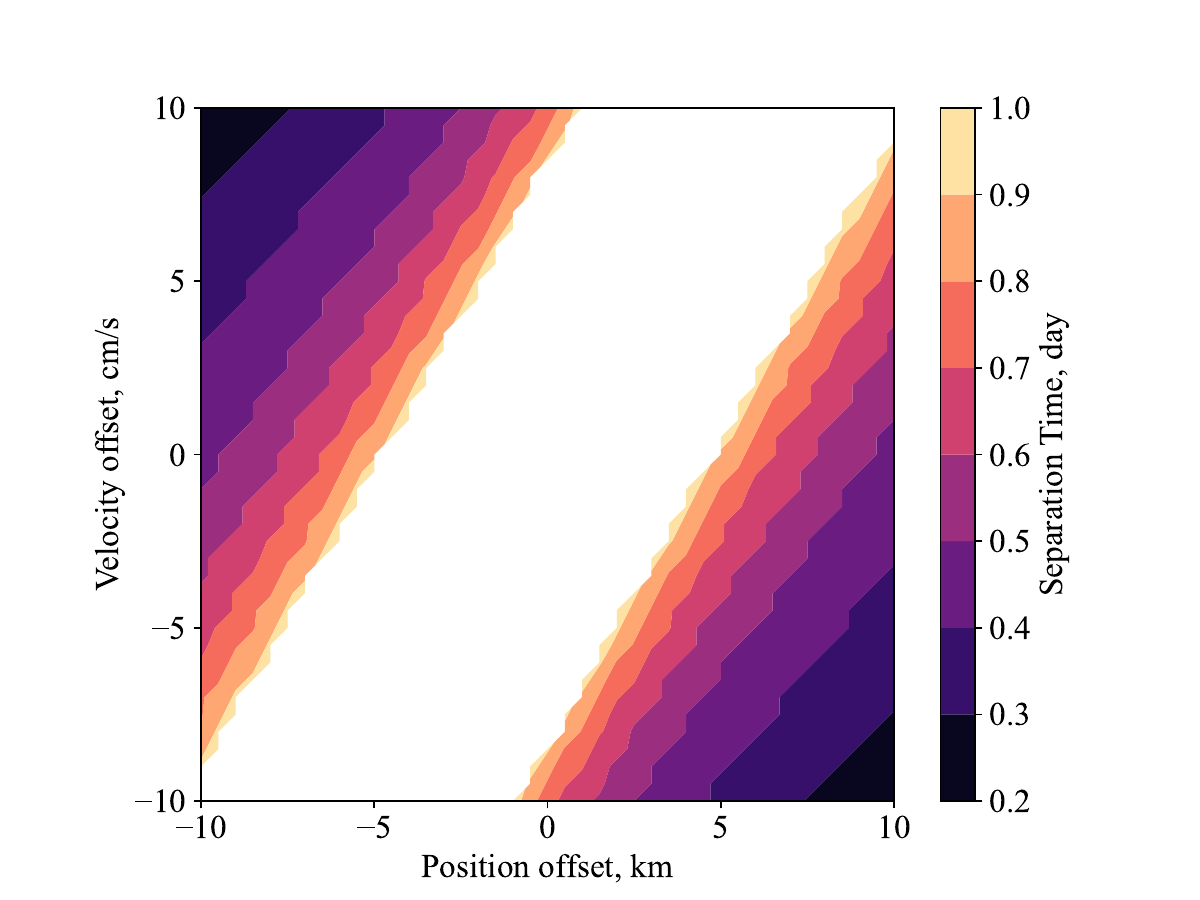}
        \caption{WSBT (a): Separation Time at time index = 0, or t=0.00 day ($T=24$ hours, stable grid)}
        \label{fig:00_tsep_graph_stable}
    \end{subfigure}

    \vspace{0.5cm} 

    \begin{subfigure}[b]{0.45\textwidth}
        \includegraphics[width=\textwidth]{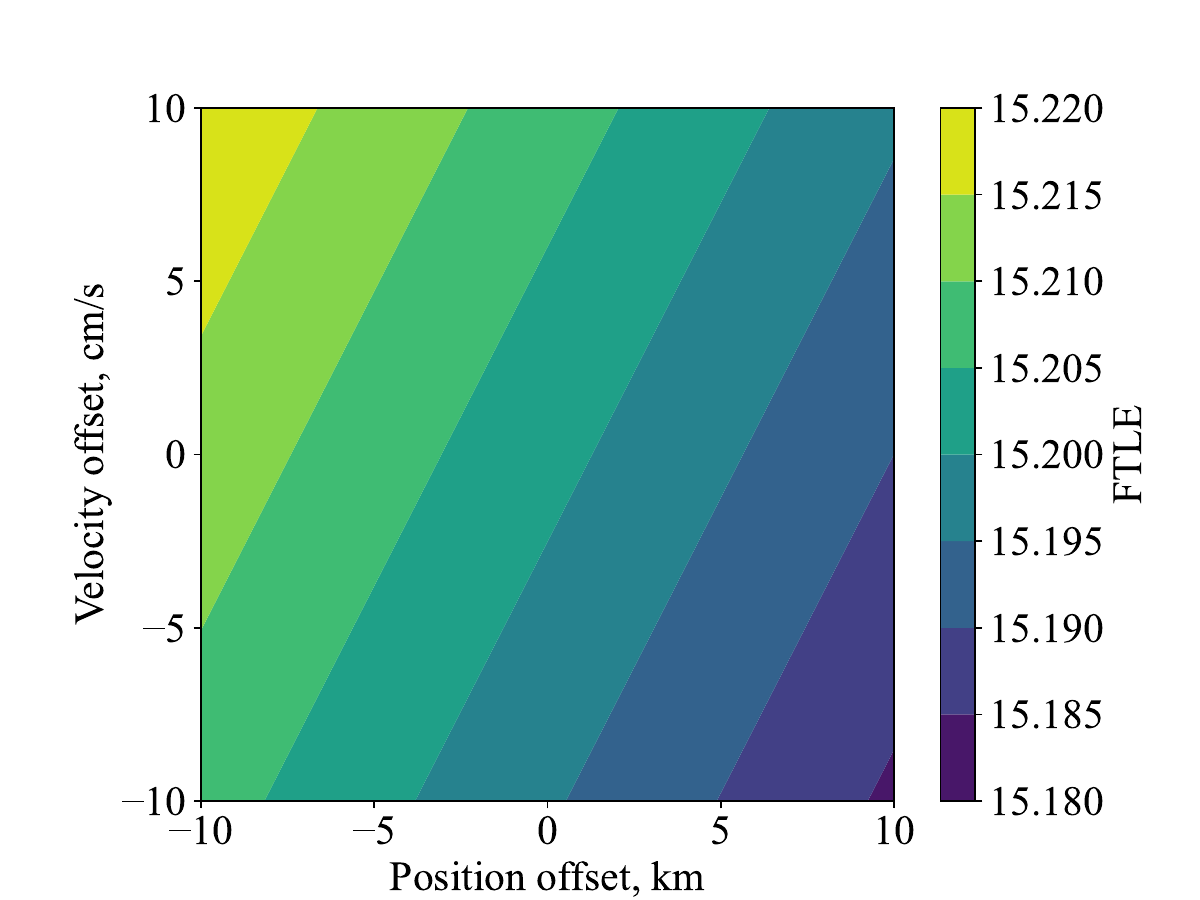}
        \caption{WSBT (a): FTLE at time index = 3, or t=4.18 day ($T=24$ hours, stable grid)}
        \label{fig:00_ftle_end_stable}
    \end{subfigure}
    \hfill
    \begin{subfigure}[b]{0.45\textwidth}
        \includegraphics[width=\textwidth]{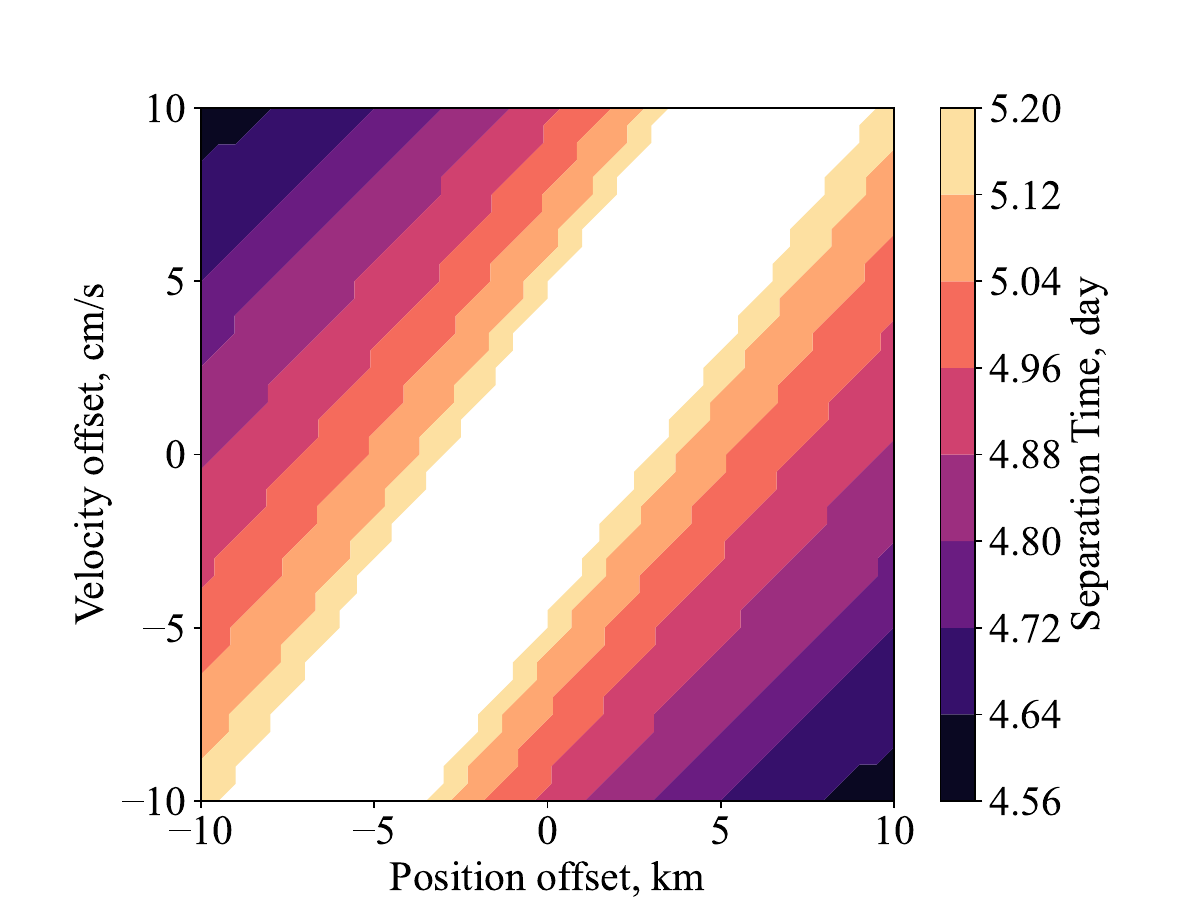}
        \caption{WSBT (a): Separation Time at time index = 3, or t=4.18 day ($T=24$ hours, stable grid)}
        \label{fig:00_tsep_graph_end_stable}
    \end{subfigure}

    \caption{Contour FTLE and Separation Time Plots for WSBT (a) with stable grid}
    \label{fig:00_contour_stable}
\end{figure}

Figure \ref{fig:00_no_tsep_rate} depicts the fraction of no separation, or the proportion of points around the nominal that remain un-diverged. This rate is plotted against points along the nominal trajectory.
Notably, in the later stages of the trajectory, specifically time indices 9-13, the fraction of no separation increases. This outcome aligns with expectations, as points situated farther from the Moon experience diminished gravitational influence and consequently exhibit greater stability. At point 2, there is noteworthy jump in the fraction of no separation which is indicative of the apolune point. At points 3 and 4 there is a drop in this fraction of no separation as the state nears its perilune point. This plot is also compared to Figure \ref{fig:00_ftle} which depicts the FTLE of the nominal state along its trajectory. While the specifc metric differs, the overall pattern shows a strong inverse correlation. 

\begin{figure}[h]
    \centering
    \includegraphics[width=0.92\linewidth]{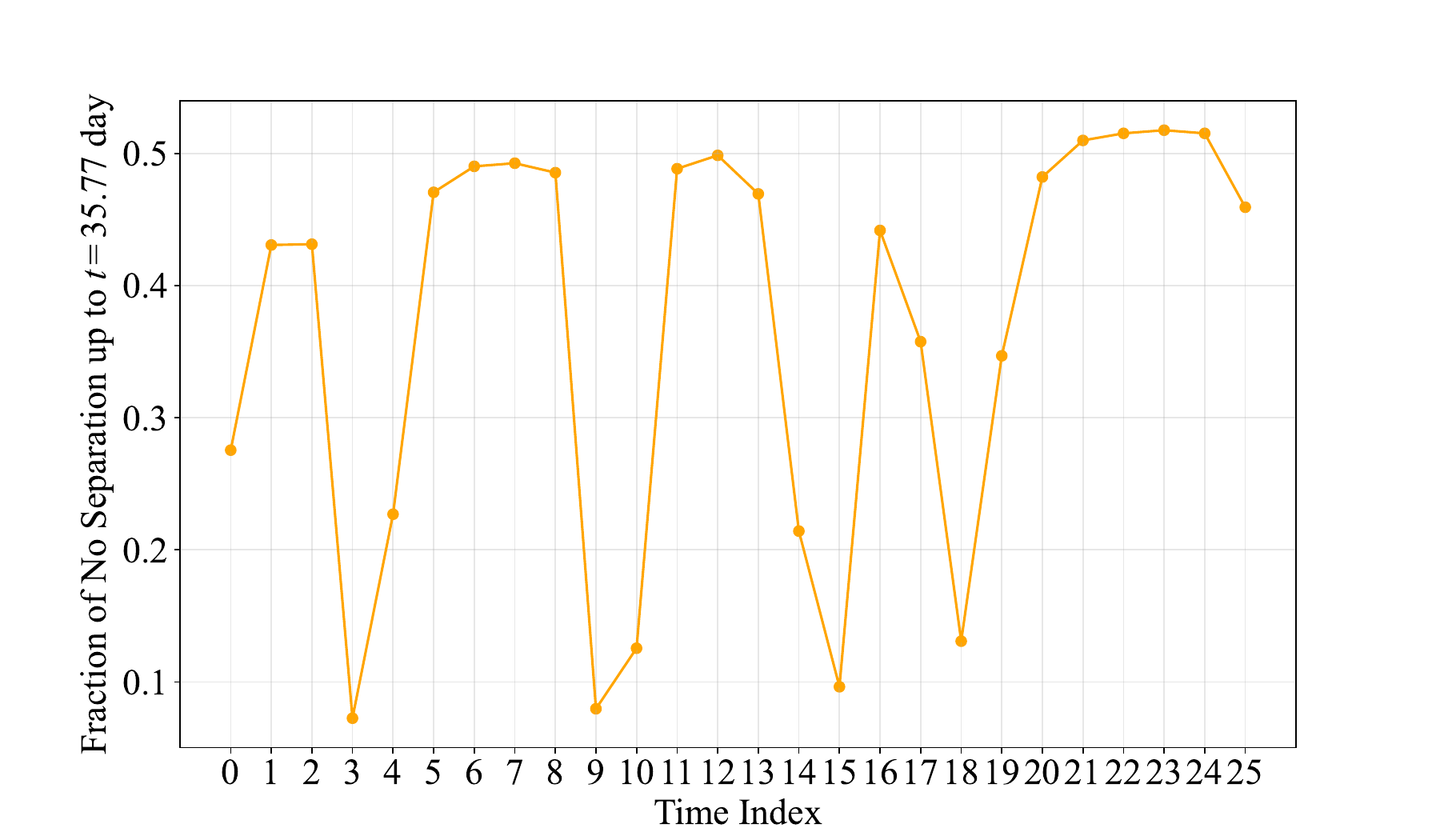}
    \caption{WSBT (b): Fraction of No Separation vs. Time Index ($T=24$ hours, stable grid, OD-ball=(10 km, 10 cm/s))}
    \label{fig:18_no_tsep_rate}
\end{figure}

\begin{figure}[h]
    \centering
    \includegraphics[width=0.92\linewidth]{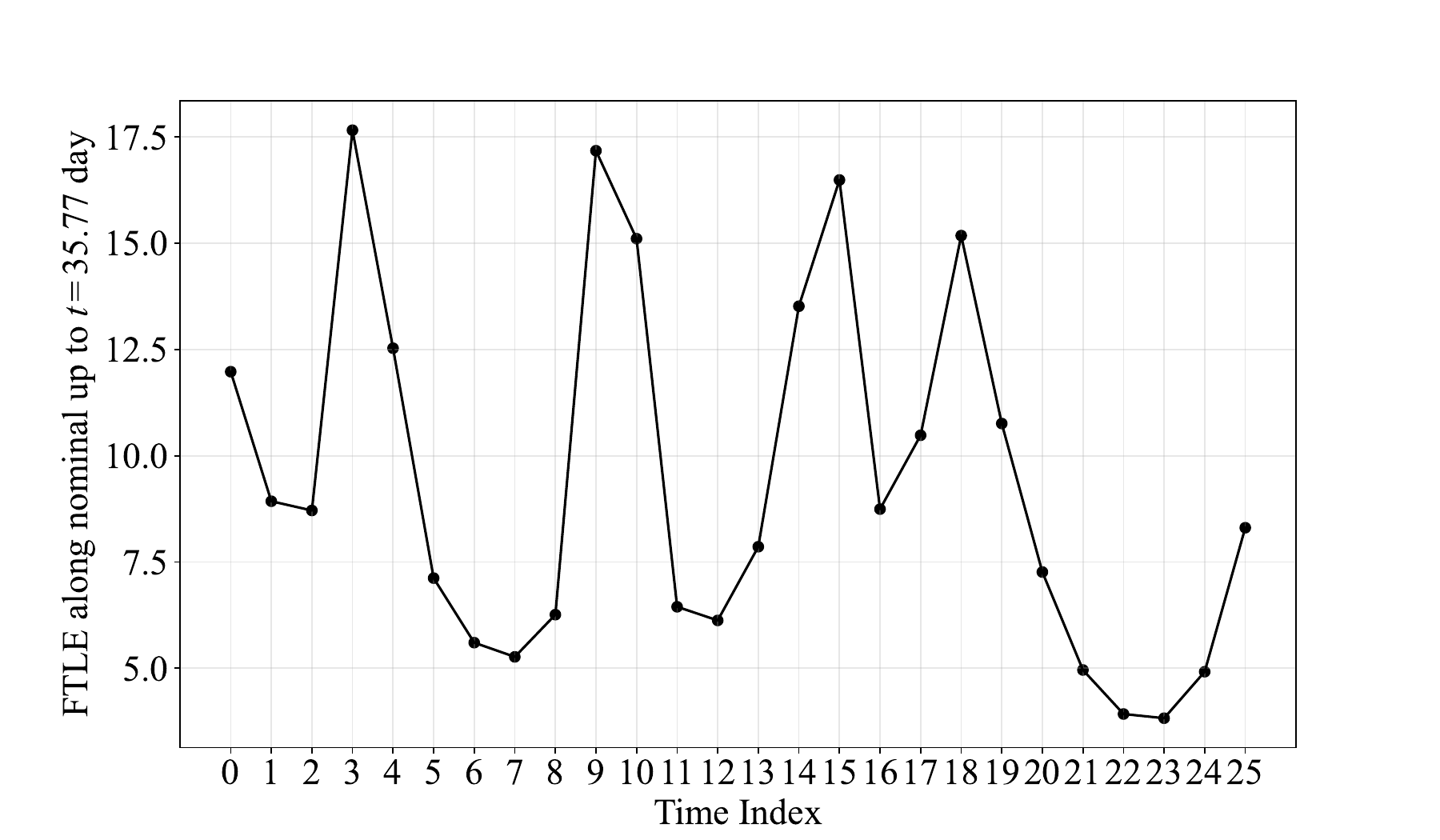}
    \caption{WSBT (b): FTLE along nominal ($T=24$ hours)}
    \label{fig:18_ftle}
\end{figure}

\begin{figure}
    \centering

    \begin{subfigure}[b]{0.45\textwidth}
        \includegraphics[width=\textwidth]{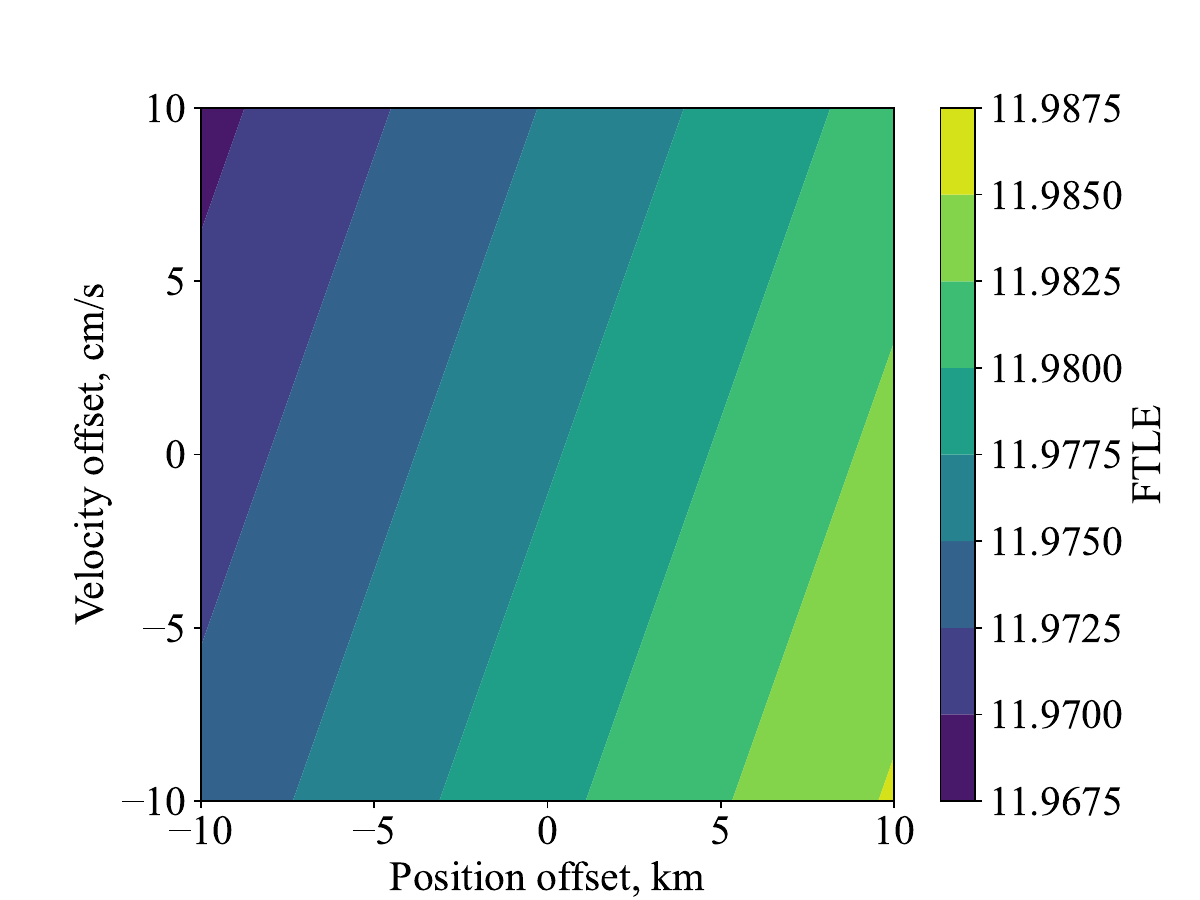}
        \caption{WSBT (b): FTLE at time index = 0, or t=0.00 day ($T=24$ hours, stable grid)}
        \label{fig:18_ftle_start_stable}
    \end{subfigure}
    \hfill
    \begin{subfigure}[b]{0.45\textwidth}
        \includegraphics[width=\textwidth]{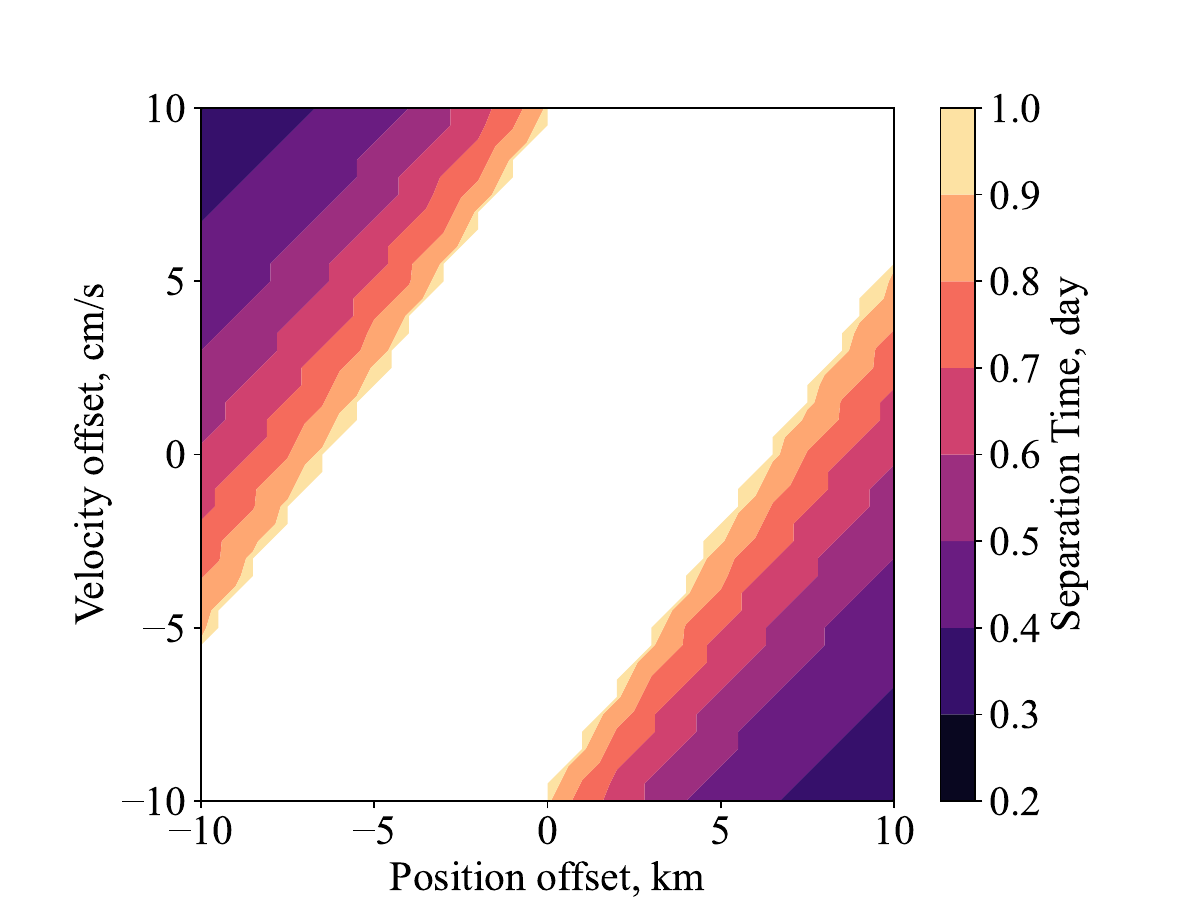}
        \caption{WSBT (b): Separation Time at time index = 0, or t=0.00 day ($T=24$ hours, stable grid)}
        \label{fig:18_tsep_graph_stable}
    \end{subfigure}

    \vspace{0.5cm} 

    \begin{subfigure}[b]{0.45\textwidth}
        \includegraphics[width=\textwidth]{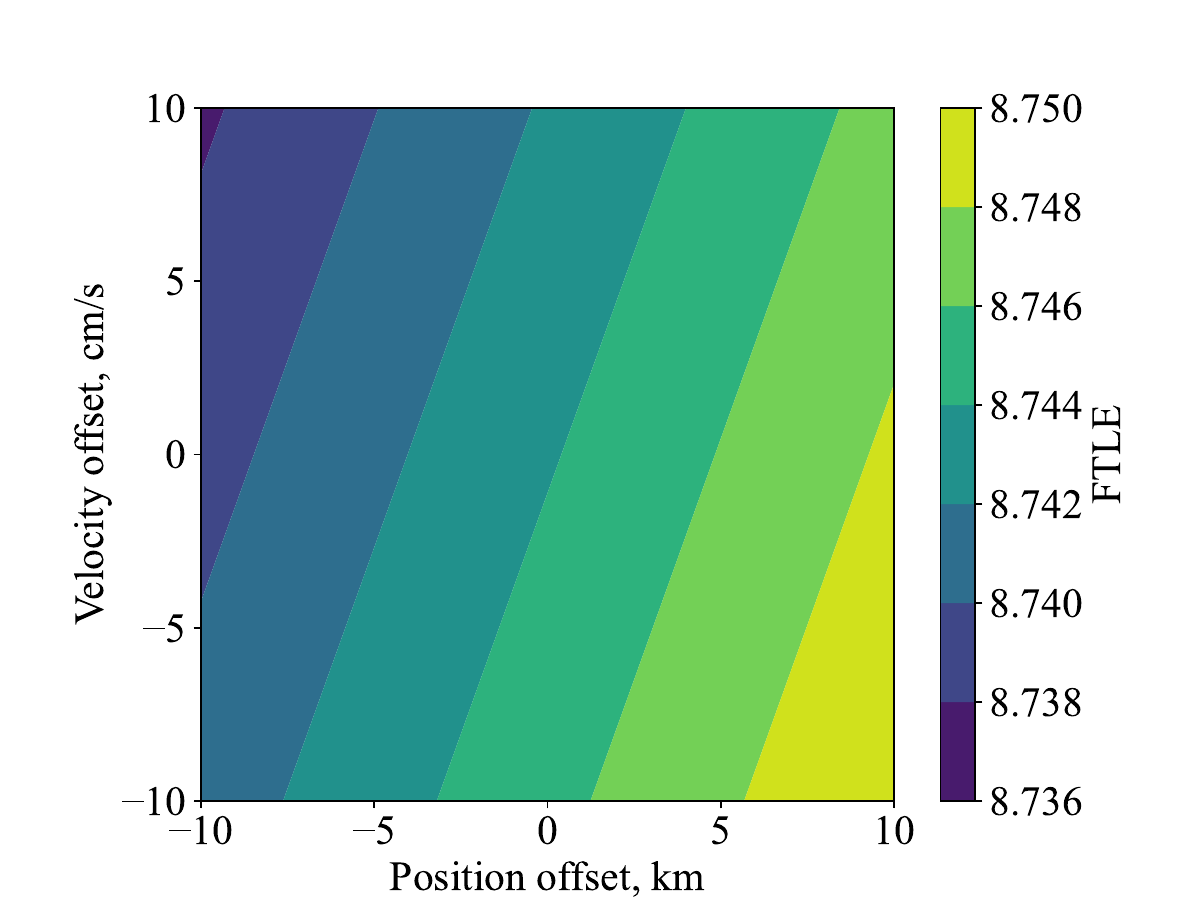}
        \caption{WSBT (b): FTLE at time index = 16, or t=22.28 day ($T=24$ hours, stable grid)}
        \label{fig:18_ftle_end_stable}
    \end{subfigure}
    \hfill
    \begin{subfigure}[b]{0.45\textwidth}
        \includegraphics[width=\textwidth]{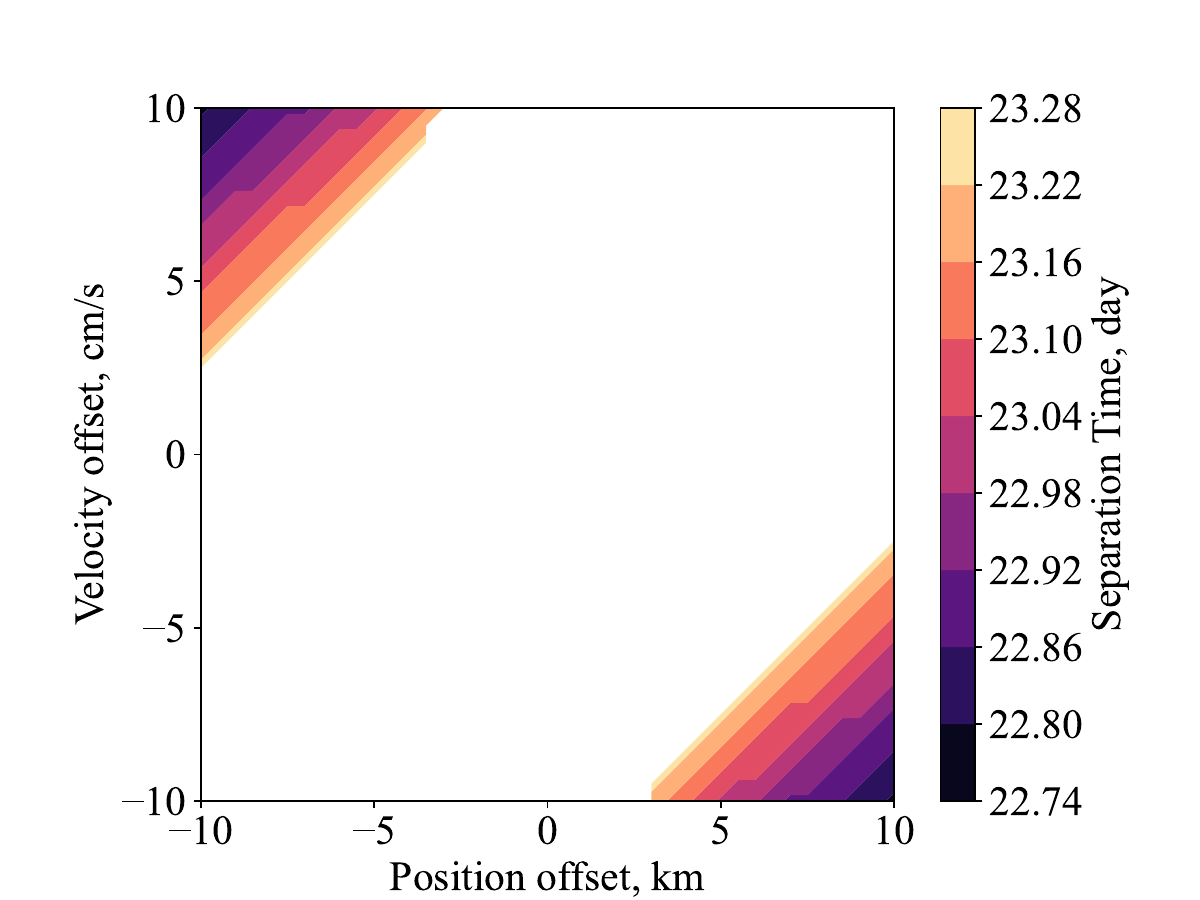}
        \caption{WSBT (b): Separation Time at time index = 16, or t=22.28 day ($T=24$ hours, stable grid)}
        \label{fig:18_tsep_graph_end_stable}
    \end{subfigure}

    \caption{Contour FTLE and Separation Time Plots for WSBT (b) with stable grid}
    \label{fig:18_contour_stable}
\end{figure}

As expected, Figure \ref{fig:18_no_tsep_rate}  displays similar trends to Figure \ref{fig:00_no_tsep_rate}. The plot is once more compared against the FTLE value of the nominal state along its trajectory in Figure \ref{fig:18_ftle}. The plots indicate that they are inverse trends of each other, indicating a strong correlation between the fraction of no separation time and the FTLE value of the nominal state. Overall, there is a relative increase in the fraction of no separation around the apolune point and areas outside of the Moon's heavy gravitational influence. The increased distance from the moon contributes to enhanced stability of the nominal WSBT, rendering SDA less favorable. This analysis is further emphasized in Figure \ref{fig:00_contour_stable} and Figure \ref{fig:18_contour_stable}, which illustrate the contour of separation time around the nominal point during its initial state and its apolune point. A comparative examination between Figure \ref{fig:00_tsep_graph_stable} and \ref{fig:00_tsep_graph_end_stable}, as well as \ref{fig:18_tsep_graph_stable} and \ref{fig:18_tsep_graph_end_stable}, reveals an increase in the time separation from its initial state to its apolune point. Once more, this underscores the heightened stability of the nominal trajectory as it extends farther from the Moon. When considering non-cooperative objects, our ability to detect them at these junctures becomes less predictable, primarily owing to the expanded range of divergence.

\section{Conclusion}
In this paper, we conducted sensitivity analysis along weak stability boundary transfers (WSBTs). The dynamical behavior studied was of a spacecraft in the ER3BP. To exploit the unpredictable nature of this system, we employed weak stability boundary transfers (WSBTs). When considering the operation of space assets, it is crucial to not only identify the nominal WSBT, but also the state space surrounding it. By examining the sensitivity of neighborhood dynamics of an object traveling along a nominal WSBT, we explore two scenarios of interest. The first scenario is concerned with the guidance and navigation of cooperative spacecraft, while the second scenario pertains to the detection of non-cooperative object.

To analyze the flow of a region in a state space, we employed the Cauchy Green Tensor (CGT) to quantify the naturally stretching directions of the flow and the finite-time Lyapunov exponent (FTLE) to quantify the flow's stretching nature. To delineate the region within the state space, we formulated a procedure for generating a grid centered around a given state on the nominal trajectory. The grid is constructed using the CGT, parameterized with respect to some time interval $T$, to identify the decoupled stretching directions in position and velocity and scaled to the chosen 6D orbit determination ball. The grid's framework facilitates the assessment of a nominal state's sensitivity. To distill a scalar metric signifying its sensitivity, we introduce the notion of separation time. The separation time, or $t_{\mathrm{sep}}$, describes the time is takes for a neighboring point to diverge a threshold distance of some 6D orbit determination ball from its nominal trajectory. To exploit the usefulness of the separation time, we selected two WSBTs to be tested alongside varying parameters pertaining to the two scenarios listed above.

In the context of guidance and navigation of cooperative spacecraft, an unstable grid was generated based on the largest eigenvector to analyze the minimum separation times along both nominal WSBTs. By examining series of plots depicting the evolution of the separation time along a nominal, FTLE contours, and separation time contours, we concluded the likelihood of a cooperative spacecraft diverging decreases when the nominal's trajectory is further from the Moon. For cooperative spacecraft, the neighboring states around a nominal's perilune point experience the greatest divergence, therefore posing challenges for executing trajectory correction maneuvers.

For the application of SDA, we evaluated the "worst-case" scenario, or areas corresponding to the slowest divergence of a neighbor state. By generating a stable grid based on the shortest eigenvector, we derived the fraction of no separation along both WSBTs, along with FTLE and separation time contours at specified points. Through analysis, we concluded the most stable areas for a nominal exist at the apolune point and the later points in its trajectory. For non-cooperative space objects, these would be the areas with the greatest uncertainty and imprecise detection.

\appendix

\bibliographystyle{AAS_publication}   
\bibliography{references}   

\end{document}